# Structural and Dynamical Behaviors of Fast Ionic Conducting Potassium *nido*-(Carba)borates

Mads B. Amdisen,*,Δ,† Hui Wu,*,† Mikael S. Andersson,δ Mirjana Dimitrievska,†,⊥,# Wei Zhou,† Torben R. Jensen,Δ Craig M. Brown,† Juscelino B. Leão,† and Terrence J. Udovic†,‡

Δ Interdisciplinary Nanoscience Center (iNANO) and Department of Chemistry, Aarhus University, 8000 Aarhus C, Denmark

†NIST Center for Neutron Research, National Institute of Standards and Technology, Gaithersburg, MD 20899-6102, United States

⊥National Renewable Energy Laboratory, Golden, CO 80401, United States

#Transport at Nanoscale Interfaces Laboratory, Swiss Federal Laboratories for Material Science and Technology (EMPA) Ueberlandstrasse 129, 8600 Duebendorf, Switzerland

δDepartment of Chemistry - Ångström Laboratory, Uppsala University, Box 538, SE-751 21 Uppsala, Sweden

‡Department of Materials Science and Engineering, University of Maryland, College Park, MD 20742, United States

ABSTRACT Solid-state batteries are one of the most recent iterations of electrochemical energy storage, and the technology can potentially provide safer and more-energy-dense batteries. The metal *closo*- and *nido*-(carba)borates show promise as versatile solid electrolytes and have been shown to have some of the highest ionic conductivities as well as wide electrochemical stability windows. In the present study, we investigate the four potassium *nido*-(carba)borates $KB_{11}H_{14}$, $K$-7-$CB_{10}H_{13}$, $K$-7,8-$C_2B_9H_{12}$, and $K$-7,9-$C_2B_9H_{12}$, and a total of eight new crystal structures were solved. All four compounds transition from a low-temperature, ordered phase to a high-temperature, disordered phase with the space group *Fm-3m*. In the high-temperature polymorphs, the anions are disordered and undergo rapid reorientational dynamics, which is confirmed by quasielastic neutron scattering experiments. Reorientational activation energies of 0.151(2) eV, 0.146(32) eV, and 0.143(3) eV were determined for $K$-7-$CB_{10}H_{13}$, $K$-7,8-$C_2B_9H_{12}$, and $K$-7,9-$C_2B_9H_{12}$, respectively. Additionally, such rotationally fluid anions are concomitant with fast potassium-ion conductivity. The highest ionic conductivity is observed for $K$-7,8-$C_2B_9H_{12}$ with $1.7 \cdot 10^{-2}$ S cm$^{-1}$ at 500 K and an activation energy of 0.28 eV in the disordered state. The differences in phase transition temperatures, reorientational dynamics, and ionic conductivities between the potassium *nido*-(carba)borates illustrate a strong correlation between the K$^+$ cationic mobility and the local cation-anion interactions, anion dynamics, and the specific positions of the carbon-atoms in the *nido*-(carba)borate anion cages.

**Introduction**

The rapidly growing demand for energy storage solutions for both mobile and stationary applications has prompted the development of technologies beyond conventional lithium-ion batteries. This is particularly apparent in the battery industry with a growing sodium-ion battery sector as well as a growing interest for solid-state batteries (SSB) due to the prospects of safer and more-energy-dense batteries.[1] While the ionic conductivity of solid-state electrolytes (SSE) is one of the major bottlenecks of SSBs,[2] recent developments in SSE research have shown that complex metal hydrides can have unparalleled ionic conductivities, even close to ambient conditions. Metal hydridoborates as SSEs have received significant attention since the discovery of the high ionic conductivity of the high-temperature polymorph of $LiBH_4$, $\sigma(Li^+) \sim 10^{-3}$ S cm$^{-1}$.[3] Since then, metal-hydridoborate-based SSEs have proliferated, and the employment of strategies such as the use of neutral ligands and hydridoborate cages with various shapes and sizes has increased in the search of an SSE that fulfills all SSB criteria.[4]

Extensive studies have revealed some of the highest ionic conductivities in the solid state for the alkali metal *closo*- and *nido*-(carba)borates, as well as wide electrochemical stability windows, making these materials some of the most promising metal hydridoborate-based SSEs.[5,6] The *closo*- and *nido*-(carba)borate anions are large polyhedral boron-hydrogen cages such as $B_{12}H_{12}^{2-}$, $CB_{11}H_{12}^-$, $B_{11}H_{14}^-$, $CB_{10}H_{13}^-$, and $C_2B_9H_{12}^-$. A selection of the *nido*-(carba)borate anion cages are visualized in Figure 1. The differences in size, shape, and charge of the anions allow for great versatility in designing SSEs with tailored properties. Examples of this are the mixed anion superionic conductors $Na_2(CB_9H_{10})(CB_{11}H_{12})$ with an ionic conductivity of $\sigma(Na^+) \sim 7 \cdot 10^{-2}$ S cm$^{-1}$ at 300 K and $0.7Li(CB_9H_{10})–0.3Li(CB_{11}H_{12})$ with an ionic conductivity of $6.7 \cdot 10^{-3}$ S cm$^{-1}$ at 298 K.[7,8] The electrochemical stability of these anions, $CB_9H_{10}^-$ and $CB_{11}H_{12}^-$, have been determined to be ~ 2.86 V vs. Li/Li$^+$ and 4.1 V vs. Na/Na$^+$, respectively.[6] While the electrochemical stability also depends on the cation, the generally high electrochemical stability of the larger borate cages, including the *nido*-borates [e.g., the electrochemical stability of $B_{11}H_{14}^-$ has been determined to be

~ 2.6 V vs. Na/Na$^+$(ref. 5)], can potentially allow the use of moderately high-voltage cathodes.[6,9] Furthermore, a Na$_4$B$_{10}$H$_{10}$B$_{12}$H$_{12}$ solid electrolyte has been reported to enable improved solid-solid interface contact and higher current density cycling.[10]

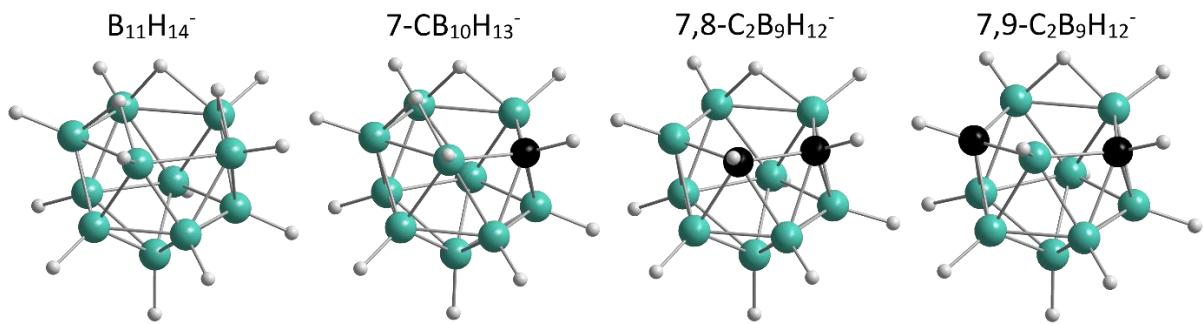

**Figure 1.** From left to right, the four different investigated *nido*-(carba)borate anions: Tetradecahydrido-*nido*-undecaborate(1-) (B$_{11}$H$_{14}^-$), tridecahydrido-*nido*-7-carbaundecaborate(1-) (7-CB$_{10}$H$_{13}^-$), dodecahydrido-*nido*-7,8-dicarbaundecaborate(1-) (7,8-C$_2$B$_9$H$_{12}^-$), and dodecahydrido-*nido*-7,9-dicarbaundecaborate(1-) (7,9-C$_2$B$_9$H$_{12}^-$). Color scheme: B (cyan), C (black), H (gray).

The high performance of the *closo*-(carba)borates has prompted investigations on their intrinsic properties to elucidate the mechanisms and effects behind the high ionic conductivities. In particular, investigations on carbon substitution of the B$_{12}$H$_{12}^{2-}$ cage have provided insights into the enhancing effects of breaking the symmetries of the borate cages.[11] Carbon substitution introduces asymmetry and an anion dipole, which affects the local anion-cation interaction and reduces the charge, which also reduces the interaction strength between anion and cation.[11] A similar effect is present in the metal *nido*-borates due to a missing atom in the icosahedral cage, which again can be further modified by the introduction of one or more carbon atoms.[12]

The high ionic conductivities are concomitant with fast anion reorientational dynamics. The fast dynamics in the *closo*- and *nido*-(carba)borates are typically introduced at elevated temperatures with a phase transition from a low-temperature ordered phase to a high-temperature disordered phase, similar to LiBH$_4$.[12–16] Because of the dipole in the carbaborate and *nido*-borate anions, the anions have preferred orientations relative to the cations, which induces frustration in the crystal structures.[11,12] Due to the otherwise high symmetry of the structures, the frustration leads to a complication of the phase transition to the low-temperature phase.[11] Consequently, this effect can support a transition to a high-temperature, highly conductive phase at lower temperatures compared to the metal *closo*-borates, and also retain the high ionic conductivity to low temperatures.[12–14] Other strategies to retain the high ionic conductivity at low temperatures, such as anion-mixing, ball-milling, and infiltration in nano-scaffold materials, have also been employed, but in most cases, these methods require a material with an already high ionic conductivity at elevated temperatures.[17–20]

In the present study, we report the properties of the four potassium *nido*-(carba)borates K-7,8-C$_2$B$_9$H$_{12}$, K-7,9-C$_2$B$_9$H$_{12}$, K-7-CB$_{10}$H$_{13}$, and KB$_{11}$H$_{14}$, including the ionic conductivities as well as the concomitant anion reorientational dynamics determined from quasielastic and inelastic neutron scattering experiments. Eight new crystal structures were determined from *in situ* synchrotron radiation powder X-ray diffraction data, all in good agreement with temperature-dependent behaviors observed from electrochemical impedance spectroscopy, differential scanning calorimetry, and quasielastic neutron scattering.

**Experimental**

Four different potassium *nido*-(carba)borates (K-7,8-C$_2$B$_9$H$_{12}$, K-7,9-C$_2$B$_9$H$_{12}$, K-7-CB$_{10}$H$_{13}$, and KB$_{11}$H$_{14}$) were obtained from Katchem.[21] The different anions are shown schematically in Figure 1. Before all measurements, any residual water that may have been present was removed by a 16 h evacuation at 443 K for K-7,8-$C_2B_9H_{12}$, 433 K for both K-7,9-$C_2B_9H_{12}$ and K-7-$CB_{10}H_{13}$, and 383 K for $KB_{11}H_{14}$.

All four samples were screened for impurities with inductively coupled plasma optical emission spectrometry (ICP-OES). The samples were dissolved in a 1% nitric acid ($HNO_3$) solution made from 67–69% PlasmaPURE[†] $HNO_3$ and Milli-Q water. All samples were prepared as ~ 100 mg $L^{-1}$ solutions to be able to detect small impurities. The high concentration entailed that the boron and potassium concentrations were outside the calibrated concentrations of the instrument, leading to some uncertainties. The measurements were carried out on a SPECTRO ARCOS inductively coupled plasma optical emission spectrometer instrument equipped with a Burgener Nebulizer and a Cyclonic Spray Chamber with an ASX-520 auto sampler.

Differential scanning calorimetry (DSC) measurements with thermogravimetric analysis (TGA) were made with a Netzsch (STA 449 *F1 Jupiter*). TGA-DSC measurements were conducted under He flow using finely ground samples in cold-weld-sealed Al sample pans. Heating/cooling rates were ±5 K·min$^{-1}$ for $KB_{11}H_{14}$ (2.0 mg), ±20 K·min$^{-1}$ for K-7-$CB_{10}H_{13}$ (9.9 mg), ±20 K·min$^{-1}$ and ±10 K·min$^{-1}$ for K-7,9-$C_2B_9H_{12}$ (3.7 mg), and ±1 K·min$^{-1}$ for K-7,8-$C_2B_9H_{12}$ (3.2 mg).

Synchrotron radiation powder X-ray diffraction (SRPXD) patterns between 298 K and 500 K were measured in sealed quartz capillaries at the Advanced Photon Source on Beamline 17-BM-B at Argonne National Laboratory [λ = 0.45399(1) Å for K-7,8-$C_2B_9H_{12}$, K-7,9-$C_2B_9H_{12}$, and K-7-CB$_{10}$H$_{13}$; λ = 0.45415(1) Å for KB$_{11}$H$_{14}$] using a two-dimensional amorphous Si-plate detector. The two-dimensional data were converted to one-dimensional data using GSAS-II.[22] Rietveld structural refinements[23] were performed using the GSAS package.[24] Additional *in situ* SRPXD patterns of K-7,8-C$_2$B$_9$H$_{12}$, K-7,9-C$_2$B$_9$H$_{12}$, and K-7-CB$_{10}$H$_{13}$ between 298 K and 500 K were measured in sealed borosilicate capillaries at MAX IV on the beamline DanMAX, λ = 0.61992 Å.

---

[†] Certain commercial equipment, instruments, or materials are identified in this document. Such identification does not imply recommendation or endorsement by the National Institute of Standards and Technology, nor does it imply that the products identified are necessarily the best available for the purpose.

All neutron scattering measurements were performed at the National Institute of Standards and Technology Center for Neutron Research using thin samples to minimize neutron beam attenuation from the highly-neutron-absorbing $^{10}$B that comprises 20 % of natural boron. Neutron vibrational spectroscopy (NVS) measurements at 4 K were done on the Filter-Analyzer Neutron Spectrometer (FANS)[25] using the Cu(220) monochromator with pre- and post-collimations of 20' of arc, yielding a full-width-at-half-maximum (fwhm) energy resolution of about 3 % of the neutron energy transfer. Neutron-elastic-scattering fixed-window scans (FWSs) within temperature limits of 225 K and 450 K were performed in heating and cooling regimens at ±0.25 K min$^{-1}$ on the High-Flux Backscattering Spectrometer (HFBS)[26], which provides a resolution of 0.8 μeV fwhm using 6.27 Å wavelength incident neutrons. Quasielastic neutron scattering (QENS) measurements up to 550 K were done on the Disc Chopper Spectrometer (DCS)[27], utilizing incident neutron wavelengths of 4.8 Å, 8 Å, 8.5 Å, and 9 Å with respective energies of 3.55 meV, 1.28 meV, 1.13 meV, and 1.01 meV; respective resolutions of 56 μeV, 30 μeV, 26 μeV, and 22 μeV fwhm; and respective maximum attainable neutron momentum transfer ($Q$) values of around 2.46 Å$^{-1}$, 1.48 Å$^{-1}$, 1.39 Å$^{-1}$, and 1.31 Å$^{-1}$. The instrumental resolution function was obtained from the purely elastic QENS measurements, typically obtained at 100 K or below. All neutron data analyses were done with the DAVE software package.[28]

Electrochemical impedance spectroscopy (EIS) data were measured with a 10 mV AC from 10 MHz to 1 Hz using a Bio-Logic MTZ-35 impedance analyzer equipped with a high-temperature sample holder. Sample pellets with a diameter of 5 mm and thicknesses between 1-2 mm were prepared at a mechanical pressure of 1 GPa. EIS data were analyzed with the software MT-Lab. The potassium ion conductivities, $\sigma(K^+)$, were calculated from the relation $\sigma = l\,R^{-1}A^{-1}$, where $l$ and $A$ are the height and area of the pellet, respectively. $R$ is the resistance of the pellet determined by fitting a Debye-type circuit to the EIS data in a Nyquist plot. The activation energies were extracted from $\ln(\sigma T)$ vs $T^{-1}$ plots.

To assist and complement the structural refinements and NVS measurements, first-principles calculations were performed within the plane-wave implementation of the generalized gradient approximation to Density Functional Theory (DFT) using a Vanderbilt-type ultrasoft potential with Perdew–Burke–Ernzerhof exchange correlation.[29] A cutoff energy of 544 eV and a 2×2×2 k-point mesh (generated using the Monkhorst-Pack scheme) were used and found to be enough for the total energy to converge within 0.01 meV/atom. All Rietveld structural refinements were performed with rigid-body anions, using their DFT energy-optimized structures. For comparison with the NVS measurements, the phonon densities of states (PDOS) were calculated for the DFT-optimized 0 K ordered structures using the supercell method with finite displacements[30,31] and were appropriately weighted to take into account the H, K, B, and C total neutron scattering cross sections.

For all figures, standard uncertainties are commensurate with the observed scatter in the data, if not explicitly designated by vertical error bars.

**Results and Discussion**

**Initial Material Characterization**

All four compounds, potassium tetradecahydrido-*nido*-7,8-dicarbaundecaborate(1-) ($KB_{11}H_{14}$), potassium tridecahydrido-*nido*-7-carbaundecaborate(1-) ($K$-7-$CB_{10}H_{13}$), potassium dodecahydrido-*nido*-7,8-dicarbaundecaborate(1-) ($K$-7,8-$C_2B_9H_{12}$), and potassium dodecahydrido-*nido*-7,8-dicarbaundecaborate(1-) (7,9-$C_2B_9H_{12}$) were characterized by nuclear magnetic resonance (NMR) spectroscopy and inductively coupled plasma optical emission spectrometry (ICP-OES) to determine their purity. $^{11}B\{^1H\}$ NMR spectra of $KB_{11}H_{14}$, $K$-7-$CB_{10}H_{13}$, $K$-7,8-$C_2B_9H_{12}$, and $K$-7,9-$C_2B_9H_{12}$ showed the expected resonances for all the anion cages with no indications of the presence of other boron-containing species; see Figure S1.[32–35] ICP-OES experiments were conducted to determine the presence of other possible mobile cations and specifically Na, but showed no signs of impurities, see Table S1.

**Thermal Behavior**

Differential scanning calorimetry (DSC) was conducted to determine the polymorphic transition behavior of each compound, see Figure 2. During heating, a polymorphic transition is observed for all compounds, reproducible at the same temperature for multiple cycles. Additionally, a second transition is observed for $K$-7,9-$C_2B_9H_{12}$ at a higher temperature. However, the signal is significantly lower compared to the first transition indicating not much enthalpic difference between these phases, see the inset in Figure 2. Notably, there is little difference in the transition temperature of the compounds during heating, regardless of heating cycle and rate. The DSC data consistently shows hysteresis, but the transitions to the low-temperature polymorphs during cooling depend on the cycling parameters. Here, we also observe a transition of $KB_{11}H_{14}$ to its low-temperature polymorph at higher temperatures than observed by Souza et al.[36], which may be due to a difference in purity. It is possible that the presence of $K_2SO_4$ in the sample made by Souza et al. has a stabilizing effect on the high-temperature polymorph.[36] Generally, if the compounds are cooled immediately after the high-temperature transition, a single relatively smooth transition to the low-temperature polymorph is observed. However, if the maximum temperature is increased slightly, or if the compound is held at the maximum temperature for an extended period, the cooling transition is shifted towards lower temperatures. Additionally, the cooling transitions become sporadic, which may be an indication of the differing thermal behaviors of various-sized crystallites in supercooled disordered states rapidly undergoing polymorphic transitions to ordered states. This possibly reflects changes in the structural quality and size of disordered-phase crystallites due to an annealing effect at elevated temperatures, which is sensitive to both time and temperature. The sporadic transitions are primarily observed for K-7-$CB_{10}H_{13}$ and K-7,8-$C_2B_9H_{12}$, see Figure **2**.

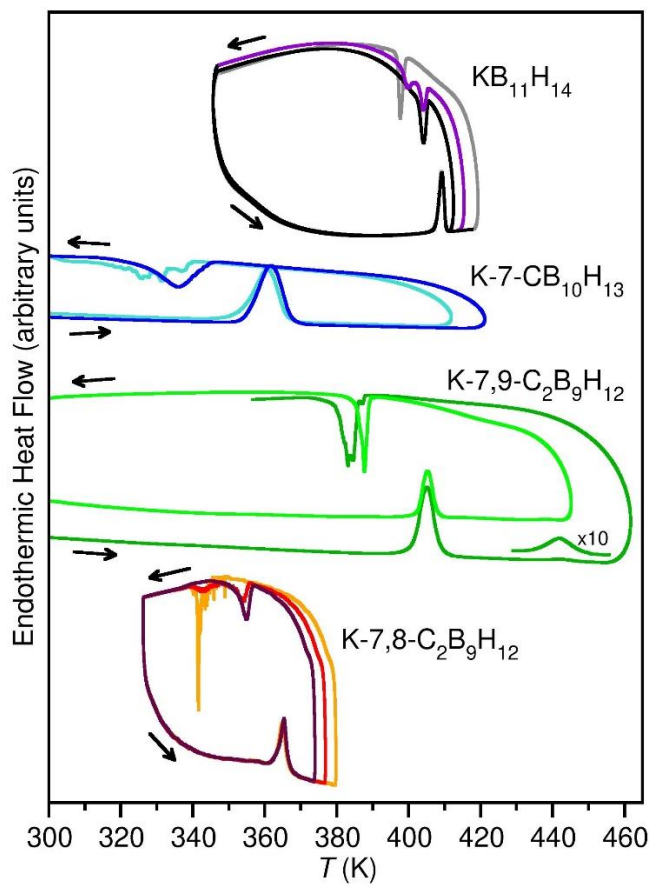

**Figure 2.** DSC data measured during heating and cooling for the potassium *nido*-(carba)borates. Data for each compound are scaled arbitrarily to facilitate the comparison between samples of different masses and heating/cooling rates. Heating and cooling segments are designated by the arrows. Color scheme: $KB_{11}H_{14}$ (black, gray, and purple), K-7-$CB_{10}H_{13}$ (blue and cyan), K-7,9-$C_2B_9H_{12}$ (dark-green (±20 K·min$^{-1}$) and light-green (±10 K·min$^{-1}$)), and K-7,8-$C_2B_9H_{12}$ (orange, red, and dark red).

**Structural Investigations**

Eight new crystal structures were determined from SRPXD data. Unit cell parameters and Rietveld refinements are provided in Table 1 and Figure S2-S10 respectively.

*Crystal Structure of Potassium Tetradecahydrido-*nido-*Undecaborate(1-)*

The low-temperature structure of potassium tetradecahydrido-*nido*-undecaborate(1-), here denoted potassium *nido*-borate, α-KB$_{11}$H$_{14}$, has previously been determined with the triclinic space group *P*-1,[36] and our results are in agreement with this structure. Each K$^+$ cation is surrounded by six B$_{11}$H$_{14}^-$ anions, forming an octahedral coordination geometry. The cation is slightly shifted from the center of the ideal octahedral position due to the dipole of the anion. With a higher positive charge at the nest-end of the anion[12,14], the anion has a preferred orientation with the nest-end pointing away from the cation. As all octahedral positions are occupied, the preferred orientation of the anion cannot be fulfilled, resulting in the cation being pushed towards one of the edges of the octahedron. The compound undergoes a first-order phase transition to its high-temperature polymorph, β-KB$_{11}$H$_{14}$, at ~ 405 K. The high-temperature polymorph has previously[36] been indexed in the space group *Fm*-3*m*, and the space group of the structure determined here is in agreement with this result; see Table 1. The cubic structure is a result of the almost isotropic disordering of the anion on the diffraction time scale and is similar to several other *fcc* metal *nido*- and *closo*-(carba)borates; see Figure 3b.[13–15,36,37]

*Crystal Structure of Potassium Tridecahydrido*-nido-*7-Carbaundecaborate(1-)*

Potassium tridecahydrido-*nido*-7-carbaundecaborate(1-), α-K-7-CB$_{10}$H$_{13}$, crystallizes in a monoclinic unit cell with space group *P*2$_1$/*a*. Each K$^+$ cation is surrounded by six 7-CB$_{10}$H$_{13}^-$ anions, forming an octahedral coordination geometry. As for the B$_{11}$H$_{14}^-$ anion, a higher localized positive charge is located at the nest-end of the 7-CB$_{10}$H$_{13}^-$ anion. The carbon atom is part of the nest-end of the anion. Carbon substitution of boron atoms in the anion cages changes the charge distribution as observed for previously calculated Mulliken charges of isolated *nido*- and *closo*-carbaborate anions.[12–14,37] Due to the higher electronegativity of carbon compared to boron, hydrogen atoms bonded to carbon have larger positive Mulliken charges relative to the rest of the hydrogen atoms in the cages.[12] This also results in a preferred orientation of the anion with the nest-end pointing away from the cation. Consequently, the cation is pushed towards one of the faces of the octahedron. α-K-7-CB$_{10}$H$_{13}$ undergoes a first-order transition to a *fcc*-type high-temperature polymorph, β-K-7-CB$_{10}$H$_{13}$, at ~ 345 K, with space group *Fm-3m*; see Figure 2 and Figure S11. The anions are described by almost isotropic disordering on the diffraction time scale, and the cations are positioned in the ideal octahedral positions.

*Crystal Structure of Potassium Dodecahydrido-nido-7,8-Dicarbaundecaborate(1-)*

Potassium dodecahydrido-*nido*-7,8-dicarbaundecaborate(1-), α-K-7,8-C$_2$B$_9$H$_{12}$, crystallizes in a monoclinic unit cell with space group *P2$_1$/n*. Similar to the two aforementioned structures, each K$^+$ cation is surrounded by six 7,8-C$_2$B$_9$H$_{12}^-$ anions, forming an octahedral coordination geometry. With an additional carbon atom, a slightly greater localized positive charge is situated on the nest-end of the 7,8-C$_2$B$_9$H$_{12}^-$ anion as compared to 7-CB$_{10}$H$_{13}^-$.[12] Again, this results in a preferred orientation of the anion with the nest-end pointing away from the cation, and the cation is pushed towards one of the faces of the octahedron; see Figure 3a. α-K-7,8-C$_2$B$_9$H$_{12}$ undergoes a first order transition to a similar *fcc*-type high-temperature polymorph, β-K-7,8-C$_2$B$_9$H$_{12}$, in space group *Fm-3m*, at ~ 360 K; see Figure S11. The anions are described by almost isotropic disordering on the diffraction time scale, and the cations are positioned in the ideal octahedral positions; see Figure 3b.

*Crystal Structure of Potassium Dodecahydrido-nido-7,9-Dicarbaundecaborate(1-)*

Potassium dodecahydrido-*nido*-7,9-dicarbaundecaborate(1-), α-K-7,9-C$_2$B$_9$H$_{12}$, crystallizes in a monoclinic unit cell with space group *P2$_1$/a*. Similar to the structures of the three aforementioned potassium *nido*-(carba)borates, each K$^+$ cation is surrounded by six 7,9-C$_2$B$_9$H$_{12}^-$ anions, forming an octahedral coordination geometry. A slightly greater localized positive charge is situated on the nest-end of the 7,9-$C_2B_9H_{12}^-$ anion compared to the 7,8-$C_2B_9H_{12}^-$, which suggests that the position and possibly the flexibility of the hydrogen atoms influences the local charge.[12] This results in a preferred orientation of the anion with the nest-end pointing away from the cation, and the cation is pushed towards one of the faces of the octahedron. Two polymorphic transitions are observed for K-7,9-$C_2B_9H_{12}$. An intermediate structure, β-K-7,9-$C_2B_9H_{12}$, is observed during the polymorphic transition from the ordered low-temperature monoclinic structure α-K-7,9-$C_2B_9H_{12}$ to the fully disordered high-temperature cubic structure γ-K-7,9-$C_2B_9H_{12}$. The polymorphic transition from α-K-7,9-$C_2B_9H_{12}$ to β-K-7,9-$C_2B_9H_{12}$ is a first-order transition occurring at ~ 395 K, whereas the transition from β-K-7,9-$C_2B_9H_{12}$ to γ-K-7,9-$C_2B_9H_{12}$ is a second-order transition occurring from ~ 450 to ~ 500 K, consistent with DSC data; see Figure S11. The intermediate β-K-7,9-$C_2B_9H_{12}$ crystallizes in a trigonal unit cell with space group *P-31c*, resembling the high-temperature structure of Na-7,8-$C_2B_9H_{12}$ (*P31c*) as well as LiCB$_9$H$_{10}$ (*P31c*).[12,13] The high-temperature γ-K-7,9-$C_2B_9H_{12}$ forms a cubic structure with the space group *Fm-3m* similar to the rest of the high-temperature potassium *nido*-(carba)borates. Disordering of the anions are observed in both β- and γ-K-7,9-$C_2B_9H_{12}$. The average structures on the diffraction time scale derived from the refinement revealed that a simple isotropic icosahedron can sufficiently describe the randomly oriented 7,9-$C_2B_9H_{12}$ anions in the β-phase while these disorderly oriented anions would be better described by an isotropic sphere in γ-phase; see Fig. 3, indicating a faster rotation of the anions at higher temperatures. In addition, the disordered anions in the β-phase arrange themselves into an *hcp* stacking (*ABAB*…) whereas they adopt an *fcc* stacking (*ABCABC*…) in the γ-phase. Such structural variations are consistent with the phase transitions observed in DSC and *in-situ* XRD, i.e., a first-order phase transition from the α- to the β-phase, which is an order-to-disorder structural transition, and a weak enthalpic second-order transition from the β- to the γ-phase, which only involves a small stacking sequence rearrangement and the rotational rate changes of anions.

The potassium *nido*-carbaborates in the present study all crystallize in similar monoclinic structures with space group $P2_1/a$ or $P2_1/n$ at room temperature. In comparison, the sodium *nido*-carbaborates are structurally slightly different with Na-7-CB$_{10}$H$_{13}$ and Na-7,9-C$_2$B$_9$H$_{12}$ crystallizing in orthorhombic unit cells, with the exception of Na-7,8-C$_2$B$_9$H$_{12}$, which also crystallizes in a monoclinic unit cell.[12] The difference in symmetry between the sodium and potassium *nido*-carbaborates suggests that the larger K$^+$ cation induces lower symmetry in the crystal structures.[12] A similar effect has been observed for the sodium and potassium *closo*-dodecacarbaborates.[14,37] The valence and the shape of the anion as well as the charge distribution over the cage are also highly influential on the symmetry of the structure and the cation coordination environment. With the icosahedral B$_{12}$H$_{12}^{2-}$ anion cages, potassium *closo*-dodecaborate crystallizes in a cubic unit cell with K$^+$ cations occupying all the tetrahedral holes, whereas the monovalent *closo*-dodecacarbaborate anion, CB$_{11}$H$_{12}^-$, entails an octahedral K$^+$ coordination environment instead, due to a dilution of the cations.[37,38] The *nido*-(carba)borate anions are generally monovalent and consequently also form structures with one cation per formula unit. Carbon substitution of boron atoms in the anion cages changes the charge distribution as observed for previously calculated Mulliken charges of isolated *nido*- and *closo*-carbaborate anions.[12–14,37] Due to the higher electronegativity of carbon compared to boron, hydrogen atoms bonded to carbon have larger positive Mulliken charges relative to the rest of the hydrogen atoms in the cages.[12] Directionality of the charge on the anions causes a preferred orientation relative to the cations, with the more positive hydrogen atoms maximizing their distance to the cations.[14,39]

**Table 1.** Crystallographic data extracted by Rietveld refinement.

| Compound | Space group | Axial lengths (Å) | Axial angles (°) | $V$ (Å$^3$) | $T$ (K)* |
|---|---|---|---|---|---|
| α-KB$_{11}$H$_{14}$ | $P$-1 | $a$ = 7.1933(4) | $α$ = 90.715(4) | 981.28(13) | 298 |
| | | $b$ = 7.0472(5) | $β$ = 85.951(4) | | |
| | | $c$ = 19.4074(9) | $γ$ = 90.009(4) | | |
| β-KB$_{11}$H$_{14}$ | $Fm$-3$m$ | $a$ = 10.1809(5) | - | 1055.24(16) | 418 |
| α-K-7-CB$_{10}$H$_{13}$ | $P2_1/a$ | $a$ = 18.8457(6) | $β$ = 91.269(3) | 949.91(8) | 298 |
| | | $b$ = 6.9700(3) | | | |
| | | $c$ = 7.2334(3) | | | |
| β-K-7-CB$_{10}$H$_{13}$ | $Fm$-3$m$ | $a$ = 10.0713(4) | - | 1021.54(12) | 415 |
| α-K-7,8-C$_2$B$_9$H$_{12}$ | $P2_1/n$ | $a$ = 10.0499(4) | $β$ = 91.0011(24) | 929.49(10) | 298 |
| | | $b$ = 12.6135(6) | | | |
| | | $c$ = 7.33361(31) | | | |
| β-K-7,8-C$_2$B$_9$H$_{12}$ | $Fm$-3$m$ | $a$ = 10.0115(10) | - | 1003.47(31) | 420 |
| α-K-7,9-C$_2$B$_9$H$_{12}$ | $P2_1/a$ | $a$ = 10.7796(4) | $β$ = 93.5186(16) | 917.97(8) | 298 |
| | | $b$ = 11.4692(4) | | | |
| | | $c$ = 7.43895(27) | | | |
| β-K-7,9-C$_2$B$_9$H$_{12}$ | $P$-31$c$ | $a$ = 6.9728(7) | - | 494.18(12) | 444 |
| | | $c$ = 11.7364(10) | | | |
| γ-K-7,9-C$_2$B$_9$H$_{12}$ | $Fm$-3$m$ | $a$ = 10.0363(4) | - | 1010.93(14) | 500 |

* Data collection temperature.

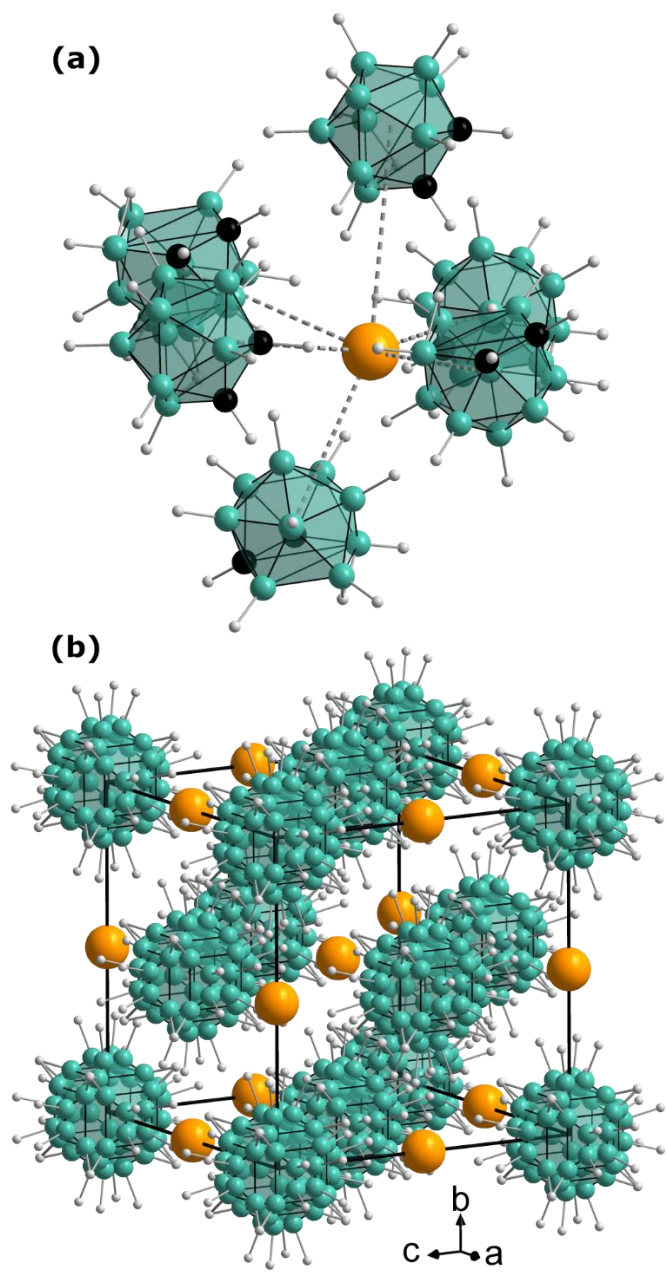

**Figure 3.** a) The octahedral coordination environment of $K^+$ in α-K-7,8-$C_2B_9H_{12}$ ($P2_1/n$). b) Representation of the high-temperature structures with space group symmetry $Fm$-$3m$. The disordered anions are a single color because C and B are indistinguishable. Color scheme: $K^+$ (orange), B (cyan), C (black), H (gray).

**Bonding Interactions**

The vibrational dynamics of the four *nido*-compounds were characterized by neutron vibrational spectroscopy (NVS) at 4 K. The neutron vibrational spectra are shown in Figure 4 in comparison with the simulated one+two-phonon densities of states of the corresponding DFT-optimized, low-temperature, ordered structures and the isolated anions. The spectrum in the measured energy region is dominated by the normal modes involving the different anion deformations. More intense vibrational features reflect normal modes with more significant H-atom displacements due to the much larger neutron scattering cross-section as well as generally larger vibrational amplitudes for H atoms compared to the other elements present. As the vibrational signature of polyhedral (carba)borate-based salt compounds is typically sensitive to the particular structural arrangement,[40] the reasonably good agreement between the experimental and simulated spectra for all four compounds corroborates the low-*T* ordered structures determined by SRPXD. The poorer agreement with the isolated anion spectra confirms the importance of lattice effects on the measured PDOS. Further information about the characters and energies of the different phonon modes contributing to the simulated PDOS for the four *nido*-compounds can be found in the animation files in the Supporting Information.[41]

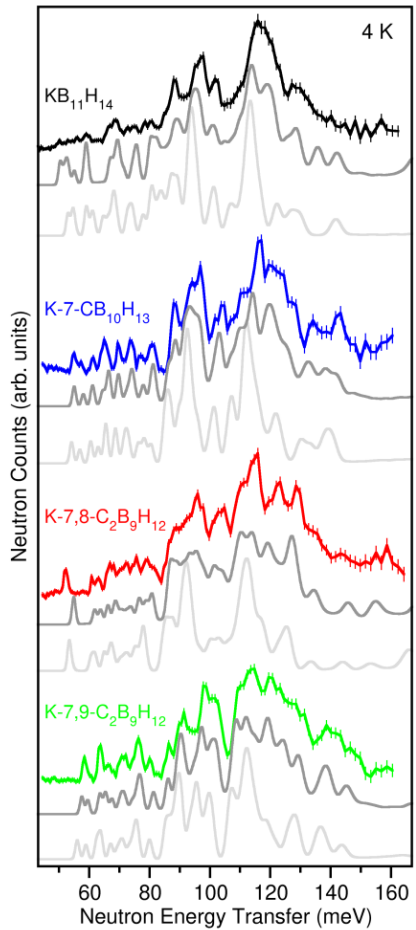

**Figure 4.** Neutron vibrational spectra of KB$_{11}$H$_{14}$ (blue), K-7-CB$_{10}$H$_{13}$ (green), K-7,8-C$_2$B$_9$H$_{12}$ (red), and K-7,9-C$_2$B$_9$H$_{12}$ (violet) at 4 K compared to the simulated one+two-phonon densities of states of the corresponding DFT-optimized, low-T, ordered structures (dark gray) and the isolated anions (light gray). Vertical error bars represent ±1σ. (N.B., 1 meV ≈ 8.0655 cm$^{-1}$.)

**Anion Reorientational Dynamics**

*Fixed Window Scans*

A change in the reorientational dynamics of the anions is also evident from a neutron elastic scattering fixed window scan (FWS). By integrating a fixed-energy slice centered around the elastic peak for each temperature, it is possible to monitor the temperature-dependent evolution of the *nido*-(carba)borate anions due to the large incoherent scattering cross-section of the hydrogen bound to the cages. When the jump frequencies increase above ~$10^8$ jumps s$^{-1}$, the elastic peak begins to broaden quasielastically, which means the intensity in the integrated region decreases. When the quasielastic part becomes so broad that it no longer contributes to the integrated region (above ~$10^{10}$ jumps s$^{-1}$), the temperature-dependent evolution of the dynamics can no longer be probed with the FWS.[12] A more detailed description of the FWS technique can be found in Ref. 42 and 43.

All four potassium *nido*-(carba)borates exhibit hysteresis, similar in magnitude to the results from DSC, of the anion dynamics based on the FWSs; see Figure 5. The transitions during heating happen within a narrow temperature range, 5-20 K, whereas the transitions during cooling are slightly slower. The FWSs of KB$_{11}$H$_{14}$ and K-7-CB$_{10}$H$_{13}$ initially decrease more rapidly than the FWSs of K-7,8-C$_2$B$_9$H$_{12}$ and K-7,9-C$_2$B$_9$H$_{12}$, suggesting that minor dynamics are introduced at lower temperatures in the two former compounds, consistent with their larger unit cell volumes at 298 K. The smaller Mulliken charges on the nest-end hemisphere of the non- and mono-carbon-substituted anions compared to the di-substituted anions may also contribute to less directionality of the anions.[12] The rapid decrease in intensity, and thereby increase in anion dynamics, of each compound during heating happens in the order K-7-CB$_{10}$H$_{13}$ (~ 345-360 K), K-7,8-C$_2$B$_9$H$_{12}$ (~ 345-370 K), K-7,9-C$_2$B$_9$H$_{12}$ (~ 395-410 K), and KB$_{11}$H$_{14}$ (~ 405-415 K); see Figure 5, consistent with the first-order phase transitions observed from DSC and *in situ* SRPXD. K-7-CB$_{10}$H$_{13}$ and K-7,8-C$_2$B$_9$H$_{12}$ show high similarity in their transition temperatures with the K-7,8-C$_2$B$_9$H$_{12}$ transition temperature range being approximately 10 K wider. A similar behavior has been observed for the equivalent sodium salts.[12] K-7,9-C$_2$B$_9$H$_{12}$ undergoes a more rapid transition than K-7,8-C$_2$B$_9$H$_{12}$, but at a 50 K higher temperature, whereas the two equivalent sodium salts reach the detection limit almost simultaneously. From *in situ* SRPXD data, β-K-7,9-$C_2B_9H_{12}$ is the dominating phase to ~ 450 K (see Figure S11), while γ-K-7,9-$C_2B_9H_{12}$ is the dominating phase at 500 K. 2 wt. % of β-K-7,9-$C_2B_9H_{12}$ persists at 500 K, which emphasizes the slow transition to fully disordered anion cages compared to the rapid transition of K-7,8-$C_2B_9H_{12}$. Additionally, the K-7,9-$C_2B_9H_{12}$ dynamics do not reach the detection limit at 410 K after the first phase transition from α-K-7,9-$C_2B_9H_{12}$ to β-K-7,9-$C_2B_9H_{12}$. However, after the initial intensity drop, a slow decrease in intensity is observed from 410 K, consistent with the second phase transition from β-K-7,9-$C_2B_9H_{12}$ to γ-K-7,9-$C_2B_9H_{12}$.

Anion dynamics for potassium *nido*-borate, $KB_{11}H_{14}$, undergo rapid changes during both heating and cooling. The anion dynamics reach the detection limit at ~ 415 K during heating and rapidly slow down again from ~ 410 K during cooling. In comparison, $NaB_{11}H_{14}$ transitions to both the disordered and the ordered state at temperatures lower than the two sodium *nido*-carbaborates, Na-7,8-$C_2B_9H_{12}$ and Na-7,9-$C_2B_9H_{12}$.[12] This suggests that the larger potassium cation favors carbon substitution in the anion cages in terms of improved rotational dynamics at lower temperatures compared to the sodium salts.

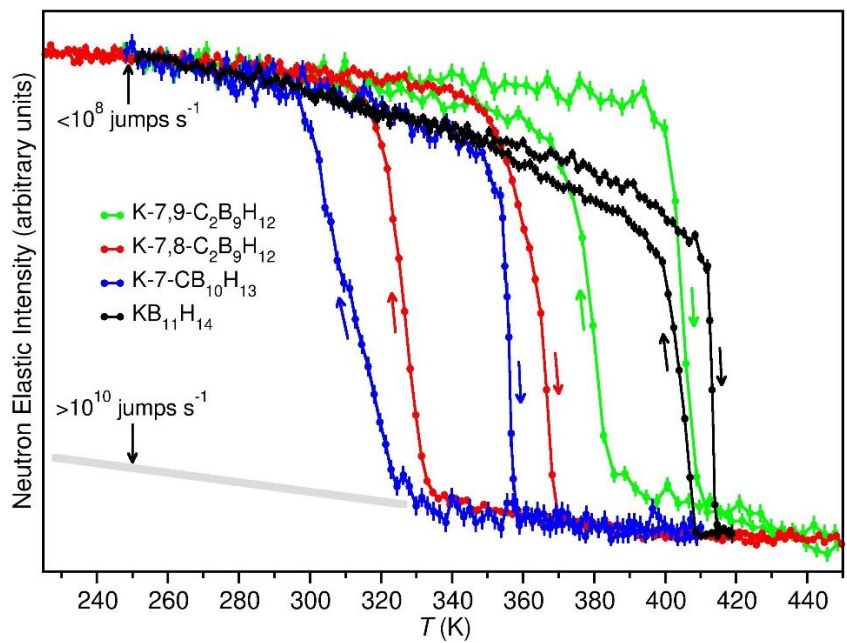

**Figure 5.** Neutron fixed window scans of the four potassium *nido*-(carba)borates. The heating and cooling directions are displayed with arrows.

*Quasielastic Neutron Scattering*

Quasielastic neutron scattering experiments were done for all four compounds in their high-temperature disordered polymorph phases, starting at the highest temperature and sequentially cooling down. Typically, one can extract jump correlation frequencies $\tau^{-1}$ associated with the Lorentzian components of the quasielastic spectra via the relation $\tau^{-1} = \Gamma/(2\hbar)$ (where $\Gamma$ is the Lorentzian full-width-half-maximum (fwhm) linewidth and $\hbar$ is the reduced Planck's constant). The temperature dependence of the fundamental jump correlation frequency $\tau_1^{-1}$ is shown in Figure 6. The $\tau_1^{-1}$ values were typically extracted from the QENS spectra using incident neutron wavelengths of 8 Å, 8.5 Å, or 9 Å at $Q = 0.8$ Å$^{-1}$, where there is minimal contamination from higher-order Lorentzian components.[10] Exemplary QENS spectra and fits are displayed in Figure S12 of the SI. All disordered polymorphs exhibit comparable $\tau_1^{-1}$ values in the range of $10^{10}$-$10^{11}$

jumps·s$^{-1}$ between 365 K and 550 K. Reorientational activation energies $E_a$ were determined from the slope ($-E_a/k_B$) of the linear fit of $\ln(\tau_1^{-1})$ vs. $T^{-1}$, where $k_B$ is the Boltzmann constant. Values of 0.151(2) eV, 0.146(32) eV, and 0.143(3) eV were determined for K-7-CB$_{10}$H$_{13}$, K-7,8-C$_2$B$_9$H$_{12}$, and K-7,9-C$_2$B$_9$H$_{12}$, respectively. We note that the value for K-7,9-C$_2$B$_9$H$_{12}$, although similar to those for the other compounds, may have been affected by the presence of the intermediate disordered β-K-7,9-C$_2$B$_9$H$_{12}$ at the lower temperatures measured. Nonetheless, no further effort was made to distinguish separate activation energies for β-K-7,9-C$_2$B$_9$H$_{12}$ and γ-K-7,9-C$_2$B$_9$H$_{12}$. The reported $E_a$ values for the nido-carbaborates are in line with 0.149(2) eV reported for the closo-carbaborate KCB$_{11}$H$_{12}$, although as seen from Figure 6, the $\tau_1^{-1}$ values for the nido-borates are somewhat less than for closo-KCB$_{11}$H$_{12}$ at the same temperatures. We note that no $E_a$ value for KB$_{11}$H$_{14}$ could be extracted since its jump correlation frequency was only measured at one temperature near its transition point ($\sim 3 \times 10^{10}$ jumps·s$^{-1}$ at 413 K), higher-temperature measurements being discouraged by the lengthy measurement time required coupled with the relatively more thermally unstable nature of this particular compound. A previous NMR study of KB$_{11}$H$_{14}$ dynamics[44] indicated a similar jump frequency and an $E_a$ value of 0.19(3) eV for the disordered β-phase in the 400 K - 418 K region. In contrast, the jump frequency for the ordered α-phase at 393 K just below the transition was found to be two-orders of magnitude lower ($\sim 10^8$ s$^{-1}$) with a substantially higher $E_a$ value of 0.53(2) eV.[44]

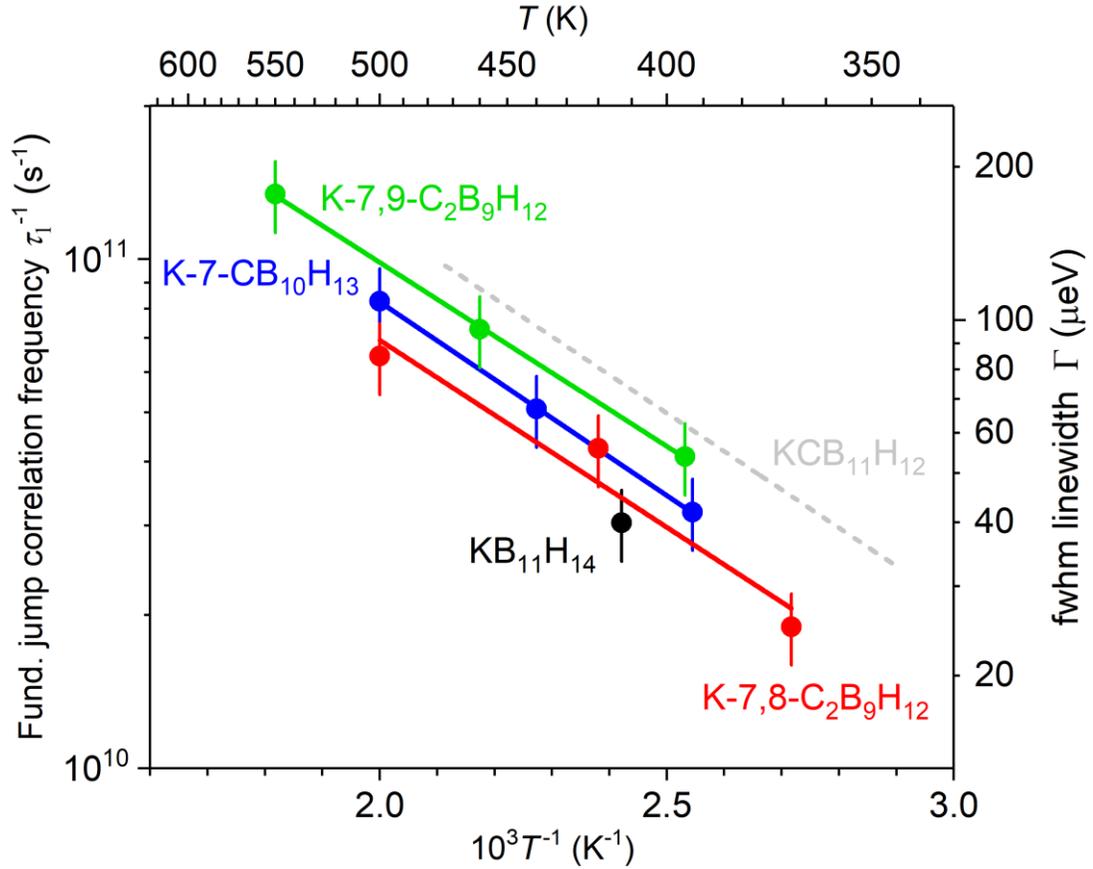

**Figure 6.** Fundamental jump correlation frequencies $\tau_1^{-1}$ of $KB_{11}H_{14}$ (black), $K$-7-$CB_{10}H_{13}$ (blue), $K$-7,8-$C_2B_9H_{12}$ (red), and $K$-7,9-$C_2B_9H_{12}$ (green) as a function of temperature compared to the behavior previously reported for $KCB_{11}H_{12}$.[32]

Further details concerning the anion reorientational mechanism in the disordered phases can be gleaned from the $Q$-dependent behavior of the elastic incoherent structure factor (EISF), which is defined as the ratio of the elastic scattering intensity to the total scattering intensity (i.e., elastic + quasielastic intensity). The experimental EISF results for $\beta$-$K$-7,8-$C_2B_9H_{12}$ at 500 K, $\gamma$-$K$-7,9-$C_2B_9H_{12}$ at 550 K, and $\beta$-$K$-7-$CB_{10}H_{13}$ at 500 K versus $Q$ are exemplified in Figure 7 and compared with calculated models for uniaxial five-fold jump reorientations, icosahedral tumbling, and both uniaxial and isotropic rotational diffusion of the anions. Unfortunately, reliable EISF data for disordered cubic β-KB$_{11}$H$_{14}$ could not be obtained from the 413 K QENS measurements, due to the presence of some substantial amount of ordered triclinic α-KB$_{11}$H$_{14}$ still persisting at this temperature. Somewhat higher temperatures are required to convert all the remaining α-KB$_{11}$H$_{14}$ to β-KB$_{11}$H$_{14}$, which were discouraged by the need for prolonged measurement times that would likely cause a gradual decomposition of the compound.

For the sake of simplicity, the uniaxial axis was chosen to be the quasi-$C_5$-symmetric axis perpendicular to the open-nest aperture of the *nido*-anions, although we have included two model curves for 7-CB$_{10}$H$_{13}$ (in gray) with a different reorientation axis passing through the carbon apex to demonstrate that reorientations around another preferred axis lead only to minor changes in the corresponding EISF and only above $Q \approx 1.5$ Å$^{-1}$.[45] Similarly, the EISFs for icosahedral tumbling and isotropic rotational diffusion are largely indistinguishable below $Q \approx 1.5$ Å$^{-1}$. The expressions representing the various model curves are included in the Supporting Information. Similarly, the EISF model differences between uniaxial five-fold jump reorientations and uniaxial rotational diffusion and between icosahedral tumbling and isotropic rotational diffusion are largely absent below $Q \approx 1.5$ Å$^{-1}$, only becoming pronounced above this $Q$ value. The expressions representing the various model curves are included in the Supporting Information.

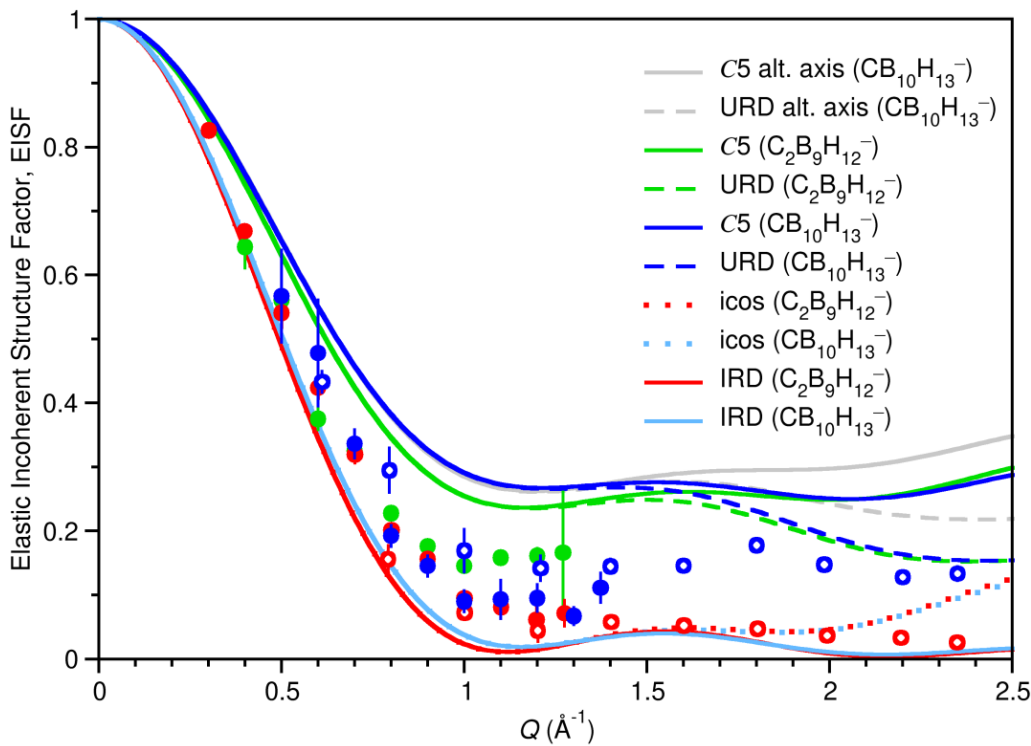

**Figure 7.** The elastic incoherent structure factors as functions of $Q$ for K-7-CB$_{10}$H$_{13}$ (500 K, blue symbols, filled: 8 Å neutrons; open: 4.8 Å neutrons), K-7,8-C$_2$B$_9$H$_{12}$ (500 K, red symbols, filled: 9 Å neutrons; open: 4.8 Å neutrons), and K-7,9-C$_2$B$_9$H$_{12}$ (550 K, green filled symbols, 9 Å neutrons) compared with calculated model curves for different reorientational mechanisms: uniaxial five-fold jumps (blue line: 7-CB$_{10}$H$_{13}$; green line: 7,8-C$_2$B$_9$H$_{12}$ and 7,9-C$_2$B$_9$H$_{12}$); uniaxial rotational diffusion (blue dashed line: K-7-CB$_{10}$H$_{13}$; green dashed line: 7,8-C$_2$B$_9$H$_{12}$ and 7,9-C$_2$B$_9$H$_{12}$); icosahedral tumbling (red dotted line: 7,8-C$_2$B$_9$H$_{12}$ and 7,9-C$_2$B$_9$H$_{12}$; light blue dotted line: 7-CB$_{10}$H$_{13}$); and isotropic rotational diffusion (red line: 7,8-C$_2$B$_9$H$_{12}$ and 7,9-C$_2$B$_9$H$_{12}$; light blue line: 7-CB$_{10}$H$_{13}$). Also included are calculated model curves for 7-CB$_{10}$H$_{13}$ assuming a different quasi-$C_5$-symmetric reorientational axis passing through the carbon apex (uniaxial five-

fold jumps: gray line; uniaxial rotational diffusion: gray dashed line). See the Supporting Information for model details.

It is apparent from Figure 7 that the EISF data for the various *nido*-compounds are already dipping below the predicted uniaxial jump model curves at 500-550 K. This suggests that the anions in these compounds at these temperatures are already reorienting themselves to various degrees around different anion axes. Similar EISF behaviors were reported for the disordered phases of *nido*-Na-7-$CB_{10}H_{13}$[45] and *closo*-$KCB_{11}H_{12}$,[37] with the latter compound results more clearly indicating a progression toward more multiaxial reorientations with increasing temperatures.

A comparison of the EISFs for the three compounds in Figure 7 suggests differences in the multidimensionality of their respective anion reorientations, with K-7,9-$C_2B_9H_{12}$ displaying behavior closer to uniaxial than K-7,8-$C_2B_9H_{12}$ and K-7-$CB_{10}H_{13}$, even though its relative temperature is 50 K higher. Compared to more intermediate behavior for K-7-$CB_{10}H_{13}$, K-7,8-$C_2B_9H_{12}$ clearly exhibits anion reorientational motions already very close to isotropic rotational diffusion, with its behavior at the highest $Q$ values much less in agreement with the EISF model curve for more-well-defined icosahedral tumbling motions. It should be noted that, despite these differences in reorientational behavior, all three anions exhibit similar fundamental jumping frequencies as displayed in Figure 6. Moreover, these rapid anion reorientational motions in the disordered phases, whether more uniaxial or highly multiaxial, are essential facilitators of increased cation translational mobility.[11]

**Potassium Ionic Conductivity**

The potassium ionic conductivities of $KB_{11}H_{14}$, $K$-7-$CB_{10}H_{13}$, $K$-7,8-$C_2B_9H_{12}$, and $K$-7,9-$C_2B_9H_{12}$ were determined from electrochemical impedance spectroscopy and are compared in Figure 8. The ionic conductivities of all four compounds are hysteretic, with the highly conducting states retained upon cooling to temperatures lower than the initial transition temperatures upon heating, consistent with the FWS data as well as the behavior of numerous other metal *closo*- and *nido*-(carba)borates.[13–16,37]

The ionic conductivities of as-synthesized and ball-milled $KB_{11}H_{14}$ have previously been reported by Souza et al. to be $\sigma(K^+) = 1.2 \cdot 10^{-4}$ S cm$^{-1}$ and $\sim 3.5 \cdot 10^{-4}$ S cm$^{-1}$, respectively, at 423 K.[36] Additionally, the two compounds have been reported to have high activation energies of $E_a$ = 1.9 eV and 1.5 eV, respectively. Here, we observe a significantly different ionic conductivity behavior of commercial $KB_{11}H_{14}$. Below 395 K, the activation energy is 1.1 eV. A distinct polymorphic transition is observed in the temperature range ~ 400 K - 410 K, consistent with the FWS data. In the high-temperature, disordered phase, the activation energy decreases to 0.38 eV, and an ionic conductivity of $\sigma(K^+) = 4.3 \cdot 10^{-4}$ S cm$^{-1}$ is reached at 423 K, comparable to the ionic conductivity of the previously reported ball-milled $KB_{11}H_{14}$.[36] The disordered structure is retained to 400 K during cooling and subsequently reverts to the ordered polymorph with low ionic conductivity.

The ionic conductivity behaviors of the three remaining compounds are also consistent with the dynamical transitions observed in the FWS data. $K$-7-$CB_{10}H_{13}$ has a distinct phase transition in the temperature range ~ 345 K - 360 K. The activation energy decreases from 1.2 eV to 0.42 eV, and an ionic conductivity of $\sigma(K^+) = 1.0 \cdot 10^{-4}$ S cm$^{-1}$ is observed at 400 K. The disordered structure is retained to 350 K during cooling before reverting to the ordered phase. Similarly, $K$-7,8-$C_2B_9H_{12}$ has a phase transition in the temperature range ~ 350 K - 370 K and a decrease in activation energy from 0.81 eV to 0.28 eV. An ionic conductivity of $1.7 \cdot 10^{-2}$ S cm$^{-1}$ is observed at 500 K, and notably K-7,8-C$_2$B$_9$H$_{12}$ is the potassium electrolyte with the highest ionic conductivity above 350 K.

The potassium ionic conductivity behavior of K-7,9-C$_2$B$_9$H$_{12}$ is slightly different from the rest of the potassium *nido*-(carba)borates, consistent with both the FWS and *in situ* SRPXD data. A distinct phase transition is observed in the ionic conductivity data in the temperature range ~ 390 K - 410 K. However, two temperature ranges with different activation energies are observed above 410 K. In the low-temperature, ordered polymorph, α-K-7,9-C$_2$B$_9$H$_{12}$, the activation energy is 1.1 eV. Following the transition to the hexagonal structure, β-K-7,9-C$_2$B$_9$H$_{12}$, the activation energy decreases to 0.55 eV. However, with the second-order phase transition to the fully disordered phase, γ-K-7,9-C$_2$B$_9$H$_{12}$, above 450 K, the activation energy decreases even further to 0.28 eV, similar to the disordered phase of K-7,8-C$_2$B$_9$H$_{12}$. An ionic conductivity of $\sigma(K^+) = 5.6 \cdot 10^{-4}$ S cm$^{-1}$ is observed at 500 K.

Notably, K-7,8-C$_2$B$_9$H$_{12}$ has an ionic conductivity more than 30 times higher than K-7,9-C$_2$B$_9$H$_{12}$. The two compounds both have an activation energy of 0.28 eV in their disordered states, which is 0.1 eV and 0.18 eV lower than the activation energies of Na-7,8-C$_2$B$_9$H$_{12}$ and Na-7,9-C$_2$B$_9$H$_{12}$, respectively. The low activation energies and especially the high ionic conductivity of K-7,8-C$_2$B$_9$H$_{12}$ supports that the higher carbon substitution in the anion cages favors mobility of the larger K$^+$ compared to Na$^+$, as Na-7,8-C$_2$B$_9$H$_{12}$ and Na-7,9-C$_2$B$_9$H$_{12}$ have the lowest ionic conductivities of the sodium *nido*-(carba)borates.[12] Na impurities in K-7,8-C$_2$B$_9$H$_{12}$ can be ruled out based on the inductively coupled plasma optical emission spectrometry results presented in Table S1. It does seem counterintuitive that disordered cubic K-7,8-C$_2$B$_9$H$_{12}$, which has the smallest unit cell volume relative to the other three disordered cubic compounds, should exhibit the highest K$^+$ ionic conductivity. This suggests that there are other factors responsible for this behavior. One possibility, which is hinted at by the QENS results, is that $K^+$ migration may be enhanced in three dimensions by the more isotropic multidimensional reorientational motions of the 7,8-$C_2B_9H_{12}^-$ anions compared to the other anions. One might expect that the isotropic reorientations of 7,8-$C_2B_9H_{12}^-$ anions in all possible directions would also better enable the migration of $K^+$ cations to different octahedral sites. In contrast, for compounds with more uniaxial anion reorientations, the $K^+$ cation motions might be more directionally restricted. It should be further noted that the nest-hydrogen on 7,8-$C_2B_9H_{12}^-$ is likely fluxional; i.e., it can rapidly move between two energetically favorable positions with a low activation barrier, which is not possible for the nest-hydrogens on 7-$CB_{10}H_{13}^-$ and 7,9-$C_2B_9H_{12}^-$. This fluxionality of the nest-hydrogen on 7,8-$C_2B_9H_{12}^-$ may be one of the reasons for the more isotropic reorientational motions of the anion, and may also flatten the energy landscape for the cation migration because of the slightly greater flexibility of the structure and the directionality of the anion charge. One caveat is that, although the three nest-hydrogens of $B_{11}H_{14}^-$ are also known to be fluxional,[46] we do not observe a similar enhancement in ionic conductivity for $KB_{11}H_{14}$ as for K-7,8-$C_2B_9H_{12}$. Based on the present experimental results, future molecular dynamics simulations as previously performed for $LiCB_{11}H_{12}$ and $NaCB_{11}H_{12}$[11] may elucidate these phenomena better.

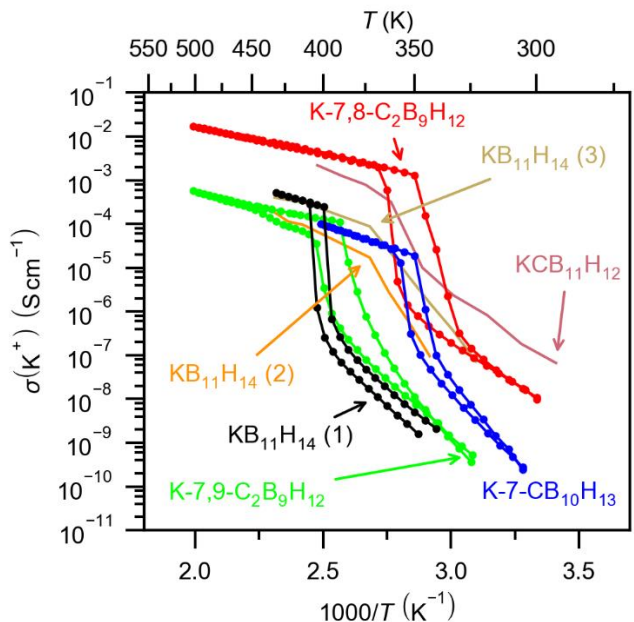

**Figure 8.** Potassium ionic conductivity data of the potassium *nido*-(carba)borates. The data on KB$_{11}$H$_{14}$ (1) are from this study, and the data on KB$_{11}$H$_{14}$ (2) and (3) are from Ref. 36, where KB$_{11}$H$_{14}$ (3) is a ball milled sample. The data on KCB$_{11}$H$_{12}$ are from Ref. 37.

**Conclusion**

This work shows that the four potassium *nido*-(carba)borates KB$_{11}$H$_{14}$, K-7-CB$_{10}$H$_{13}$, K-7,8-C$_2$B$_9$H$_{12}$, and K-7,9-C$_2$B$_9$H$_{12}$ all undergo first-order polymorphic transitions from low-temperature, ordered structures to high-temperature, disordered structures with high ionic conductivities, similar to other metal *closo*- and *nido*-(carba)borates. For all four compounds, the disordered high temperature polymorphs have cubic structures, where the anions are orientationally disordered. Additionally, K-7,9-C$_2$B$_9$H$_{12}$ has an intermediate phase with a trigonal unit cell in the temperature range ~ 450 K to ~ 500 K. There is good agreement among the DSC data, the observed crystallographic phase transitions, the quasielastic neutron scattering fixed window scans, and the ionic conductivity data. The fixed window scans show that the hydrogen jumps in all four compounds increase from ~$10^8$ jumps s$^{-1}$ to at least the probing limit of ~$10^{10}$ jumps s$^{-1}$. Consequently, the high ionic conductivities are concomitant with fast reorientational dynamics of the anion cages. The reorientational activation energies of the anions in K-7-CB$_{10}$H$_{13}$, K-7,8-C$_2$B$_9$H$_{12}$, and K-7,9-C$_2$B$_9$H$_{12}$ were determined to be 0.151(2) eV, 0.146(32) eV, and 0.143(3) eV, respectively. The ionic conductivity activation energies in the high-temperature phases are 0.38 eV, 0.42 eV, 0.28 eV, and 0.28 eV for KB$_{11}$H$_{14}$, K-7-CB$_{10}$H$_{13}$, K-7,8-C$_2$B$_9$H$_{12}$, and K-7,9-C$_2$B$_9$H$_{12}$, respectively. Notably, these activation energies are similar to or lower than the equivalent sodium *nido*-(carba)borates.[12] While there is a clear trend in the ionic conductivities of the sodium *nido*-(carba)borates, i.e. lower ionic conductivity with increasing carbon substitution in the anion cage, there is no clear trend for the potassium *nido*-(carba)borates. The highest ionic conductivity is observed for K-7,8-C$_2$B$_9$H$_{12}$, which has an ionic conductivity of $1.7 \cdot 10^{-2}$ S cm$^{-1}$ at 500 K, which is more than 30 times higher than the ionic conductivity of K-7,9-C$_2$B$_9$H$_{12}$ at the same temperature. Both K-7-CB$_{10}$H$_{13}$ and K-7,8-C$_2$B$_9$H$_{12}$ transition to their high-temperature polymorphs at significantly lower temperatures than KB$_{11}$H$_{14}$ and K-7,9-C$_2$B$_9$H$_{12}$. This shows that the ionic conductivity is highly dependent on the cation, the local anion charge and structure, the local interactions between the cation and the various anion cages, and the influence of anion reorientational behavior. Furthermore, relatively small changes in the anion has a large impact on the cation conductivity as seen when comparing the conductivities of K-7,8-C$_2$B$_9$H$_{12}$ and K-7,9-C$_2$B$_9$H$_{12}$. The present experimental study lays the groundwork for future molecular dynamics studies, which may elucidate the influence of the size and charge density of the cation as well as the local anion charge and structure on the local interactions in the crystal structures. This may also provide a better understanding of the ionic conduction mechanisms, which can facilitate better rational design of future metal hydridoborate-based solid-state electrolytes.

The promising results of especially K-7,8-$C_2B_9H_{12}$ suggests the possibility of stabilizing the high ionic conductivity to lower temperatures via strategies such as anion-mixing,[5,7] ball-milling,[19,47] nano-scaffold infiltration,[18] or complexation with molecular ligands.[48] With K-7,8-$C_2B_9H_{12}$ being one of the better solid-state potassium-ion conductors, there are prospects of the compound being a relevant electrolyte for solid-state potassium batteries. However, further electrochemical studies are necessary to elucidate the electrochemical stability of the electrolytes as well as the compatibility with various cathodes and anodes. Through the realization of an electrochemically stable electrolyte with a high ionic conductivity at low temperatures, the development of a practical solid-state potassium battery operational at room temperature may be possible.

## ASSOCIATED CONTENT

**Supporting Information**.

Nuclear magnetic resonance spectroscopy data, inductively coupled plasma optical emission spectrometry data, Rietveld refinements, *in situ* synchrotron radiation powder X-ray diffraction data, quasielastic neutron scattering results, ionic conductivity activation energy results (PDF). Phonon animation files of $KB_{11}H_{14}$, K-7-$CB_{10}H_{13}$, K-7,8-$C_2B_9H_{12}$, and K-7,9-$C_2B_9H_{12}$ (zip).

## AUTHOR INFORMATION

**Corresponding Author**

*Mads B. Amdisen - Interdisciplinary Nanoscience Center (iNANO) and Department of Chemistry, Aarhus University, Langelandsgade 140, 8000 Aarhus C, Denmark; Email: mba@inano.au.dk

*Hui Wu - NIST Center for Neutron Research, National Institute of Standards and Technology, Gaithersburg, MD 20899-6102, United States; Email: hui.wu@nist.gov

**Author Contributions**

The manuscript was written through contributions of all authors. All authors have given approval to the final version of the manuscript.

ACKNOWLEDGMENT

The work was supported by the Danish Council for Independent Research, Nature and Universe (Danscatt), and Technology and Production (CaMBat, DFF 0217-00327B). Affiliation with the Center for Integrated Materials Research (iMAT) at Aarhus University is gratefully acknowledged. Funding from the Danish Ministry of Higher Education and Science through the SMART Lighthouse is gratefully acknowledged. The authors gratefully acknowledge Dr. B. A. Trump for assistance with the SRPXD measurements. M.S.A. acknowledges the support from the ÅForsk Foundation (21-453) and the Göran Gustafsson Foundation. We acknowledge the MAX IV Laboratory for beamtime on the DanMAX beamline under proposal 20230301. Research conducted at MAX IV, a Swedish national user fa cility, is supported by Vetenskapsrådet (Swedish Research Council, VR) under contract 2018-07152, Vinnova (Swedish Governmental Agency for Innovation Systems) under contract 2018-04969 and Formas under contract 2019-02496. DanMAX is funded by the NUFI grant no. 4059-00009B. Access to the HFBS was provided by the Center for High Resolution Neutron Scattering, a partnership between the National Institute of Standards and Technology and the National Science Foundation under Agreement DMR-2010792. Powder X-ray diffraction experiments were performed on Beamline 17-BM-B at the Advanced Photon source, a U.S. Department of Energy Office of Science User Facility operated by Argonne

National Laboratory, supported by the U.S. Department of Energy Office of Basic Energy Sciences (DE-AC02-06CH11357).

# SUPPORTING INFORMATION

Structural and Dynamical Behaviors of Fast Ionic Conducting Potassium *nido*-(Carba)borates

Mads B. Amdisen,*,Δ,† Hui Wu,*,† Mikael S. Andersson,δ Mirjana Dimitrievska,†,⊥,# Wei Zhou,† Torben R. Jensen,Δ Craig M. Brown,† Juscelino B. Leão,† and Terrence J. Udovic†,‡

Δ Interdisciplinary Nanoscience Center (iNANO) and Department of Chemistry, Aarhus University, 8000 Aarhus C, Denmark

†NIST Center for Neutron Research, National Institute of Standards and Technology, Gaithersburg, MD 20899-6102, United States

⊥National Renewable Energy Laboratory, Golden, CO 80401, United States

#Transport at Nanoscale Interfaces Laboratory, Swiss Federal Laboratories for Material Science and Technology (EMPA) Ueberlandstrasse 129, 8600 Duebendorf, Switzerland

δDepartment of Chemistry - Ångström Laboratory, Uppsala University, Box 538, SE-751 21 Uppsala, Sweden

‡Department of Materials Science and Engineering, University of Maryland, College Park, MD 20742, United States

*Authors to whom correspondence should be addressed. E-mail: mba@inano.au.dk; hui.wu@nist.gov

**Nuclear magnetic resonance spectroscopy**

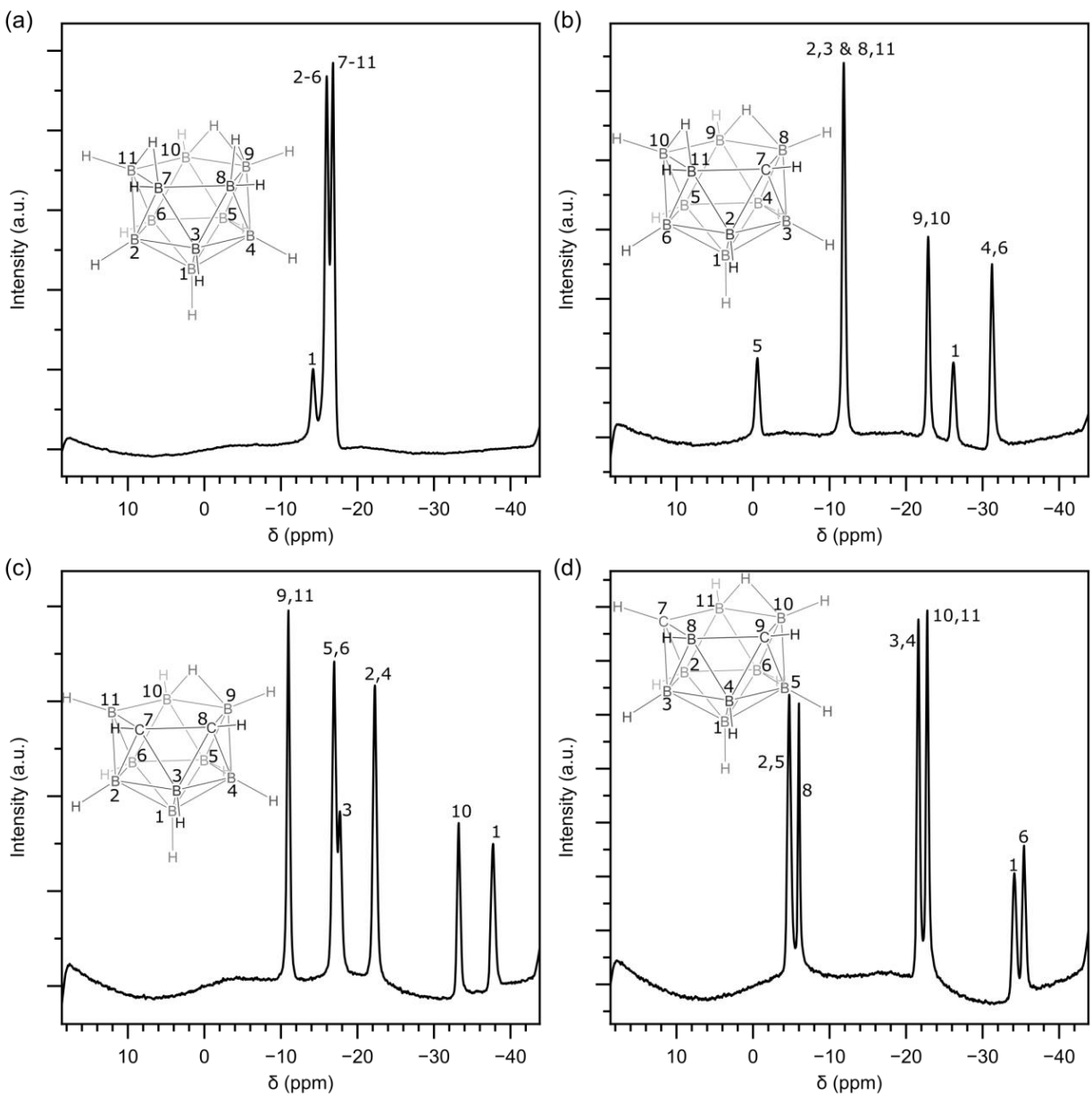

**Figure S1.** $^{11}B\{^1H\}$ nuclear magnetic resonance spectra of a) $KB_{11}H_{14}$, b) $K$-7-$CB_{10}H_{13}$, c) $K$-7,8-$C_2B_9H_{12}$, and d) $K$-7,9-$C_2B_9H_{12}$.

**Inductively coupled plasma optical emission spectrometry**

**Table S1.** Results from inductively coupled plasma optical emission spectrometry. A semi-quantitative method was used. The results have been normalized to the boron content. The theoretical K/B values were calculated under the assumption of the samples being completely pure.

| Sample | B/B | K/B theo. | K/B exp. | Na/B | Cl/B* |
|---|---|---|---|---|---|
| $KB_{11}H_{14}$ | 1 | 0.32880 | 0.33935 | - | 0.00115 |
| $K\text{-}7\text{-}CB_{10}H_{13}$ | 1 | 0.36168 | 0.36432 | - | 0.00357 |
| $K\text{-}7,8\text{-}C_2B_9H_{12}$ | 1 | 0.40187 | 0.42522 | - | 0.00632 |
| $K\text{-}7,9\text{-}C_2B_9H_{12}$ | 1 | 0.40187 | 0.45434 | - | 0.00684 |

* Part of the Cl signal is due to an impurity in the solvent (not quantified).

**Rietveld refinements**

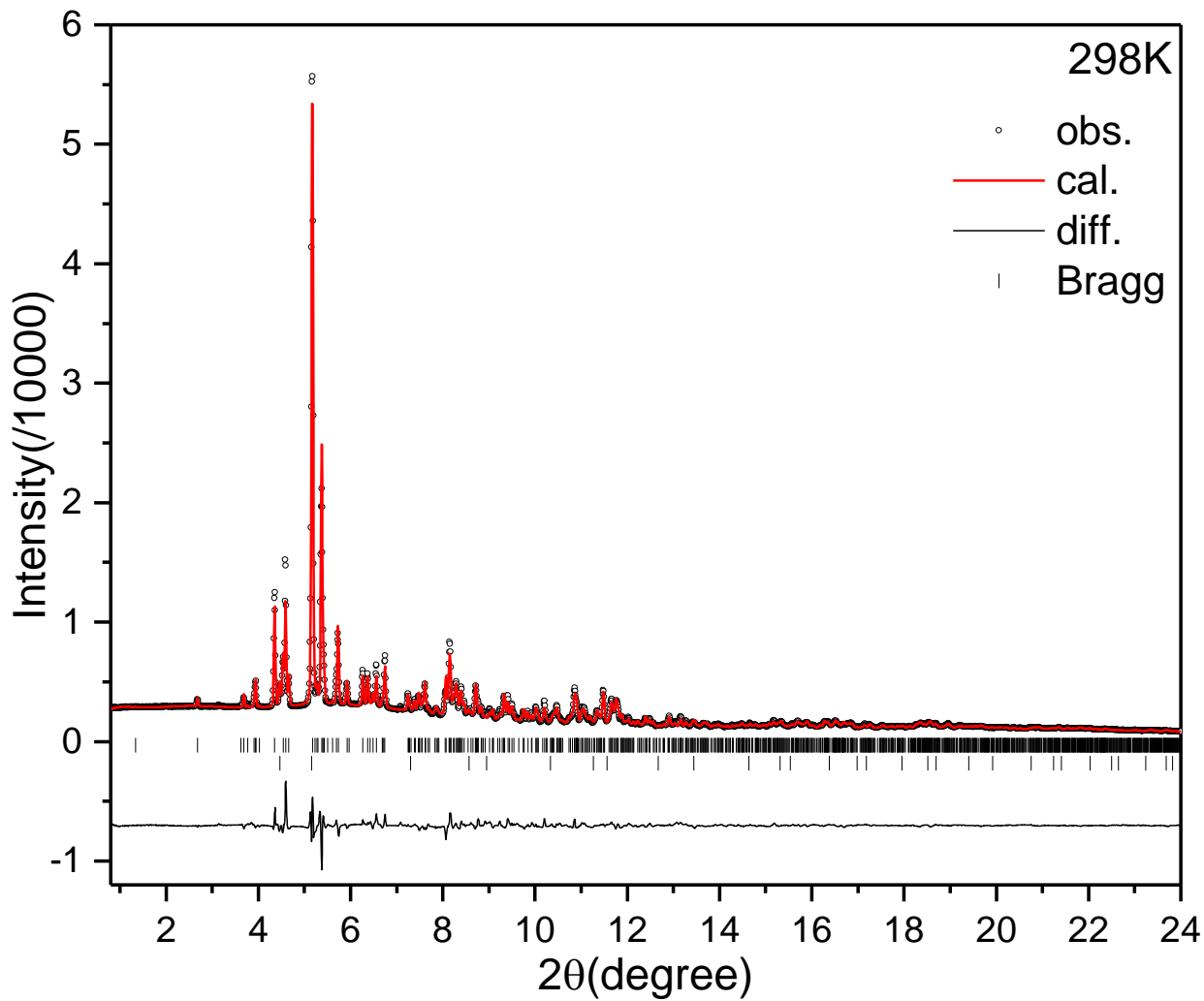

**Figure S2.** Experimental (circles), fitted (line), and difference (line below observed and calculated patterns) XRD profiles for $KB_{11}H_{14}$ at 298 K ($\lambda=0.45415$Å). Vertical bars indicate the calculated positions of Bragg peaks of $KB_{11}H_{14}$ (S.G. *P*-1, wt Frac. 91.41(3) %) and disordered HT phase (S.G. *Fm*-3*m*, wt Frac. 8.6(3) %) (from the top). $R_{wp}=0.0537$, $R_p=0.0371$, $\chi^2=2.56$. Refined lattice parameters of triclinic phase $KB_{11}H_{14}$ at room temperature: *a*=7.1933(4)Å, *b*=7.0472(5) Å, *c*=19.4074(9) Å, α=90.715(4)°, β=85.951(4)°, γ=90.009(4)°, and V=981.28(13)Å$^3$.

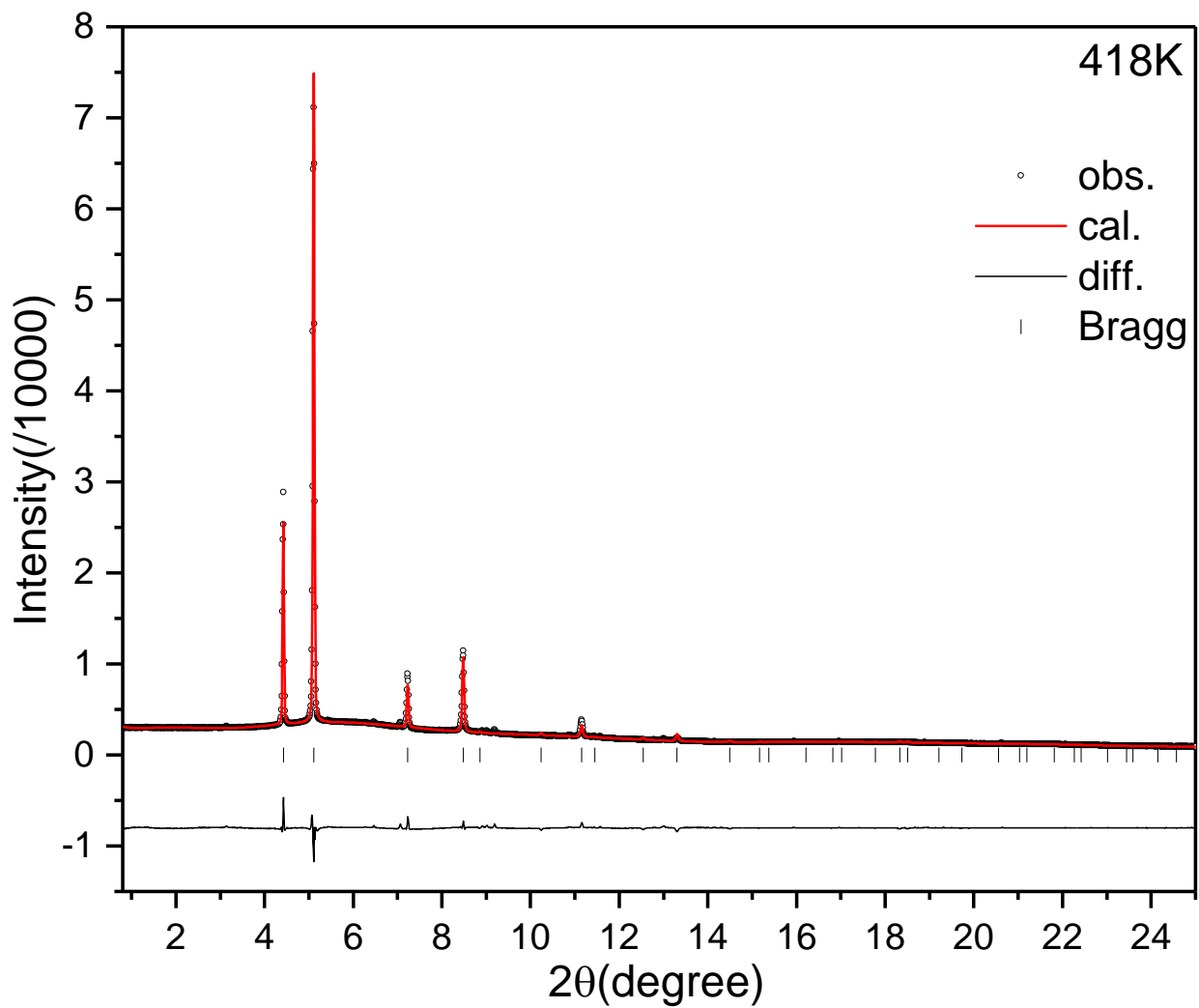

**Figure S3.** Experimental (circles), fitted (line), and difference (line below observed and calculated patterns) XRD profiles for $KB_{11}H_{14}$ at 418 K ($\lambda=0.45415$Å). Vertical bars indicate the calculated positions of Bragg peaks of $KB_{11}H_{14}$ (S.G. *Fm-3m*). $R_{wp}$=0.0312, $R_p$=0.0206, $\chi^2$=1.48. Refined lattice parameters of high-temperature disordered phase $KB_{11}H_{14}$: *a*=10.1809(5)Å and V=1055.24(16)Å$^3$.

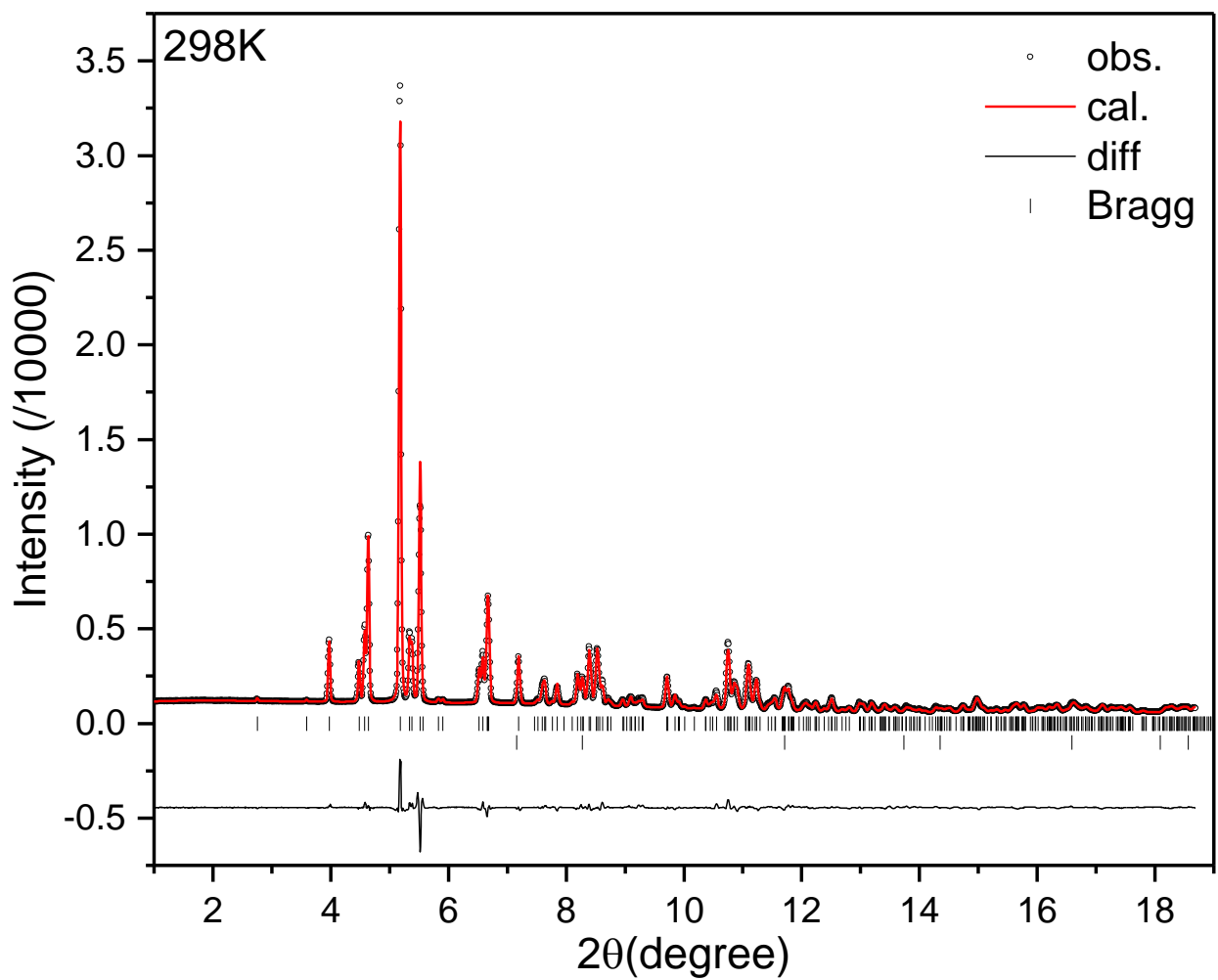

**Figure S4.** Experimental (circles), fitted (line), and difference (line below observed and calculated patterns) XRD profiles for $KCB_{10}H_{13}$ at 298 K ($\lambda$=0.45399Å). Vertical bars indicate the calculated positions of Bragg peaks of $KCB_{10}H_{13}$ (S.G. $P2_1/a$, Wt. Frac.: 99.844(2) %) and KCl (Wt. Frac.: 0.16(2) %) (from the top). $R_{wp}$=0.0409, $R_p$=0.0265, $\chi^2$=1.47. Refined lattice parameters of the room-temperature $KCB_{10}H_{13}$ phase: $a$=18.8457(6)Å, $b$=6.9700(3) Å, $c$=7.2334(3) Å, $\beta$=91.269(3)°, and V=949.91(8)Å$^3$.

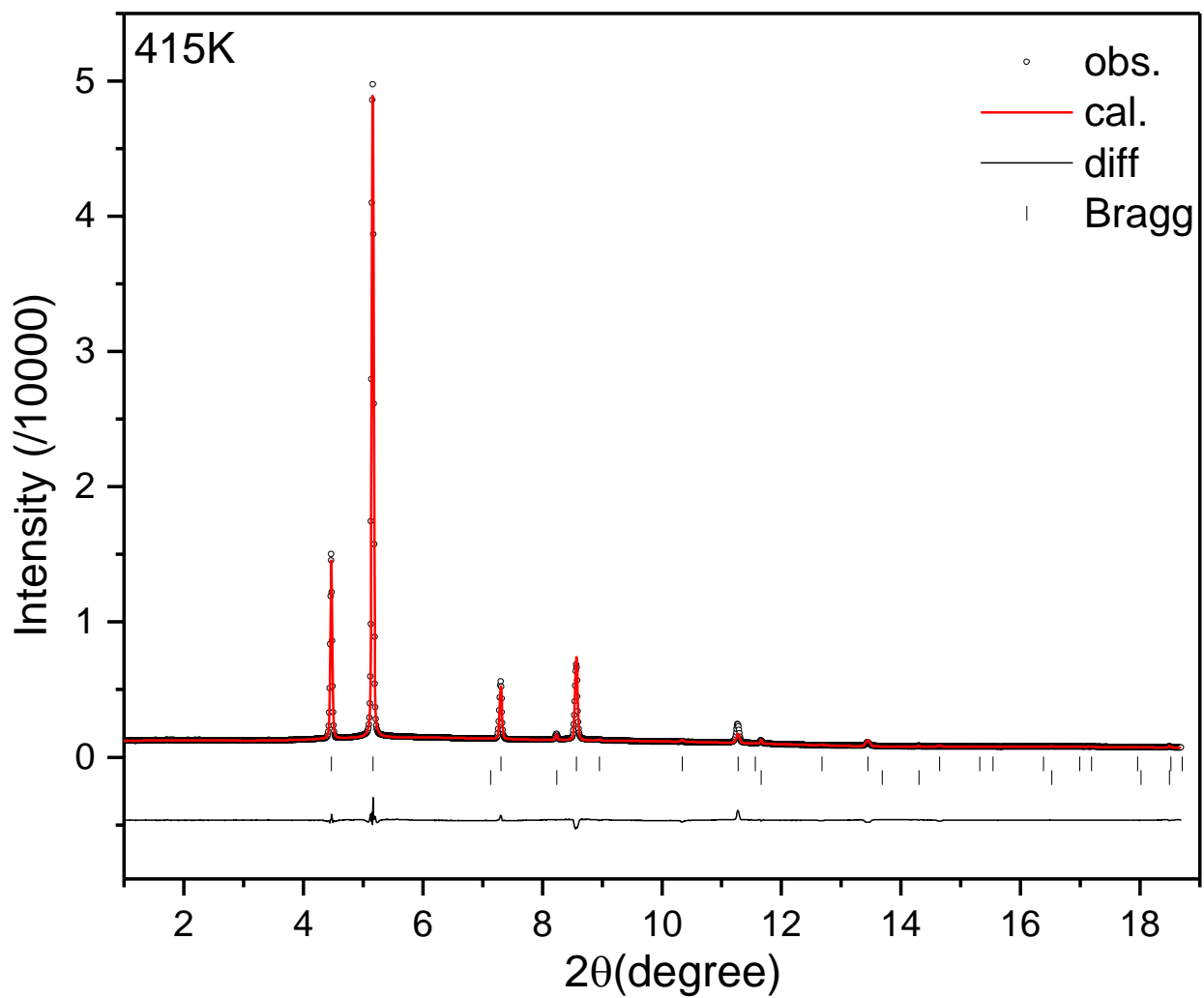

**Figure S5.** Experimental (circles), fitted (line), and difference (line below observed and calculated patterns) XRD profiles for $KCB_{10}H_{13}$ at 415 K ($\lambda$=0.45399Å). Vertical bars indicate the calculated positions of Bragg peaks of $KCB_{10}H_{13}$ (S.G. *Fm-3m*, Wt. Frac.: 99.800(4) %) and KCl (Wt. Frac.: 0.200(8) %) (from the top). $R_{wp}$=0.0332, $R_p$=0.0196, $\chi^2$=1.19. Refined lattice parameters of the high-temperature $KCB_{10}H_{13}$ phase: *a*=10.0713(4)Å and V=1021.54(12)Å$^3$.

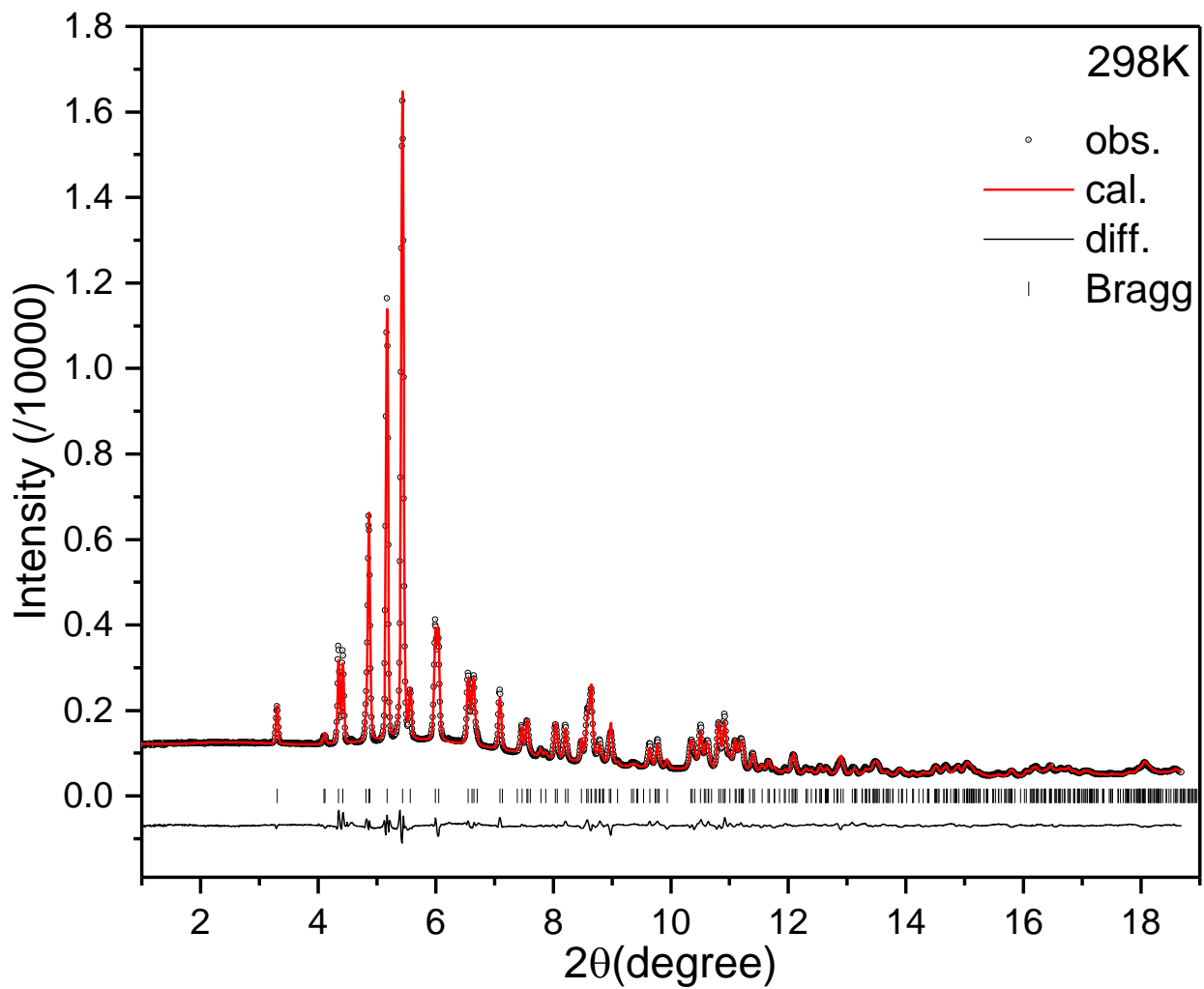

**Figure S6.** Experimental (circles), fitted (line), and difference (line below observed and calculated patterns) XRD profiles for K-7,8-$C_2B_9H_{12}$ at 298 K ($\lambda$=0.45399Å). Vertical bars indicate the calculated positions of Bragg peaks of K-7,8-$C_2B_9H_{12}$ (S.G. $P2_1/n$). $R_{wp}$=0.0300, $R_p$=0.0219, $\chi^2$=0.99. Refined lattice parameters of room-temperature phase K-7,8-$C_2B_9H_{12}$: $a$=10.0499(4)Å, $b$=12.6135(6) Å, $c$=7.33361(31) Å, $\beta$=91.0011(24)°, and V=929.49(10)Å$^3$.

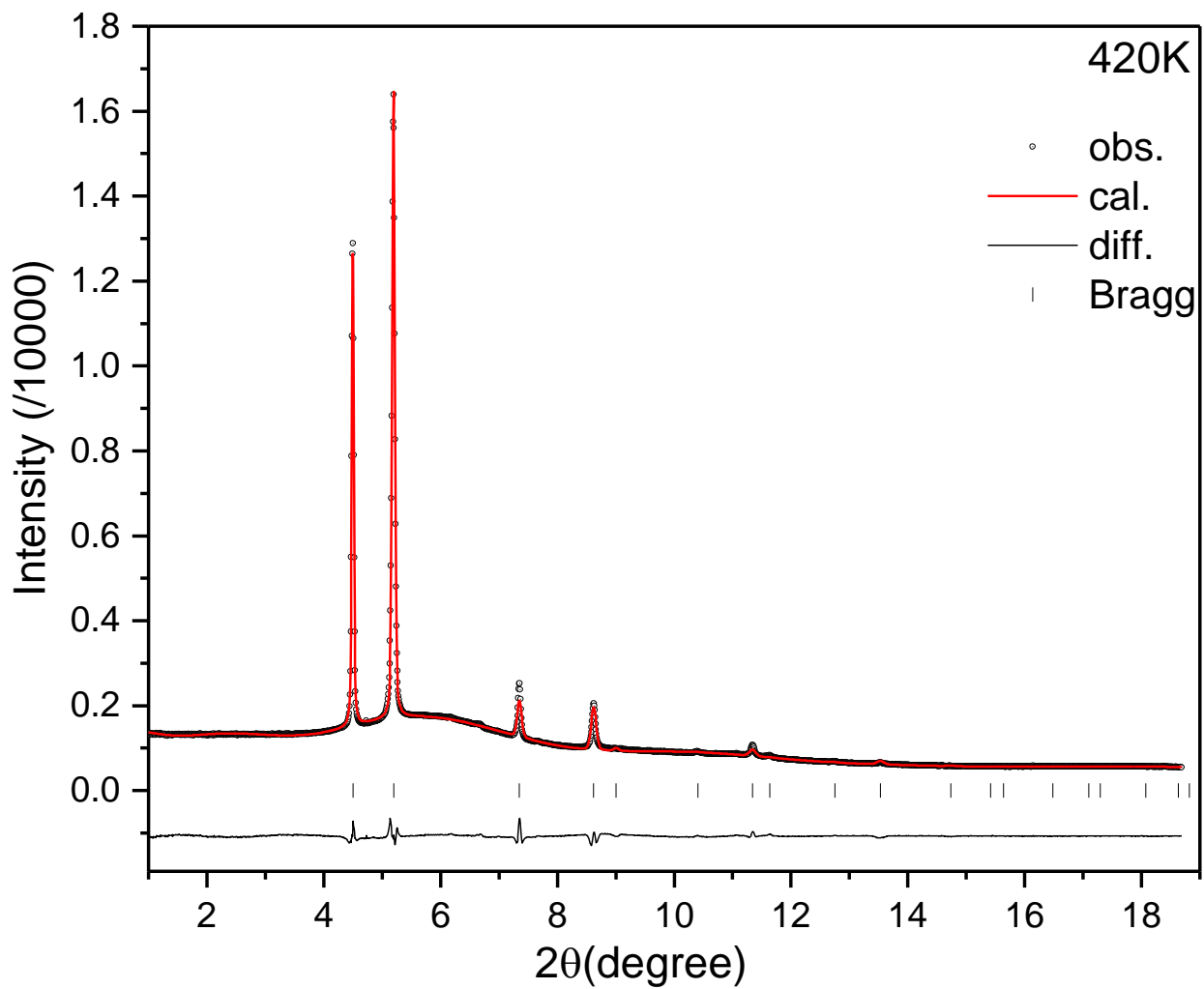

**Figure S7.** Experimental (circles), fitted (line), and difference (line below observed and calculated patterns) XRD profiles for K-7,8-$C_2B_9H_{12}$ at 420 K ($\lambda$=0.45399Å). Vertical bars indicate the calculated positions of Bragg peaks of K-7,8-$C_2B_9H_{12}$ (S.G. *Fm-3m*). $R_{wp}$=0.0260, $R_p$=0.0167, $\chi^2$=0.86. Refined lattice parameters of high-temperature phase $KCB_{10}H_{13}$: *a*=10.0115(10)Å and V=1003.47(31)Å$^3$.

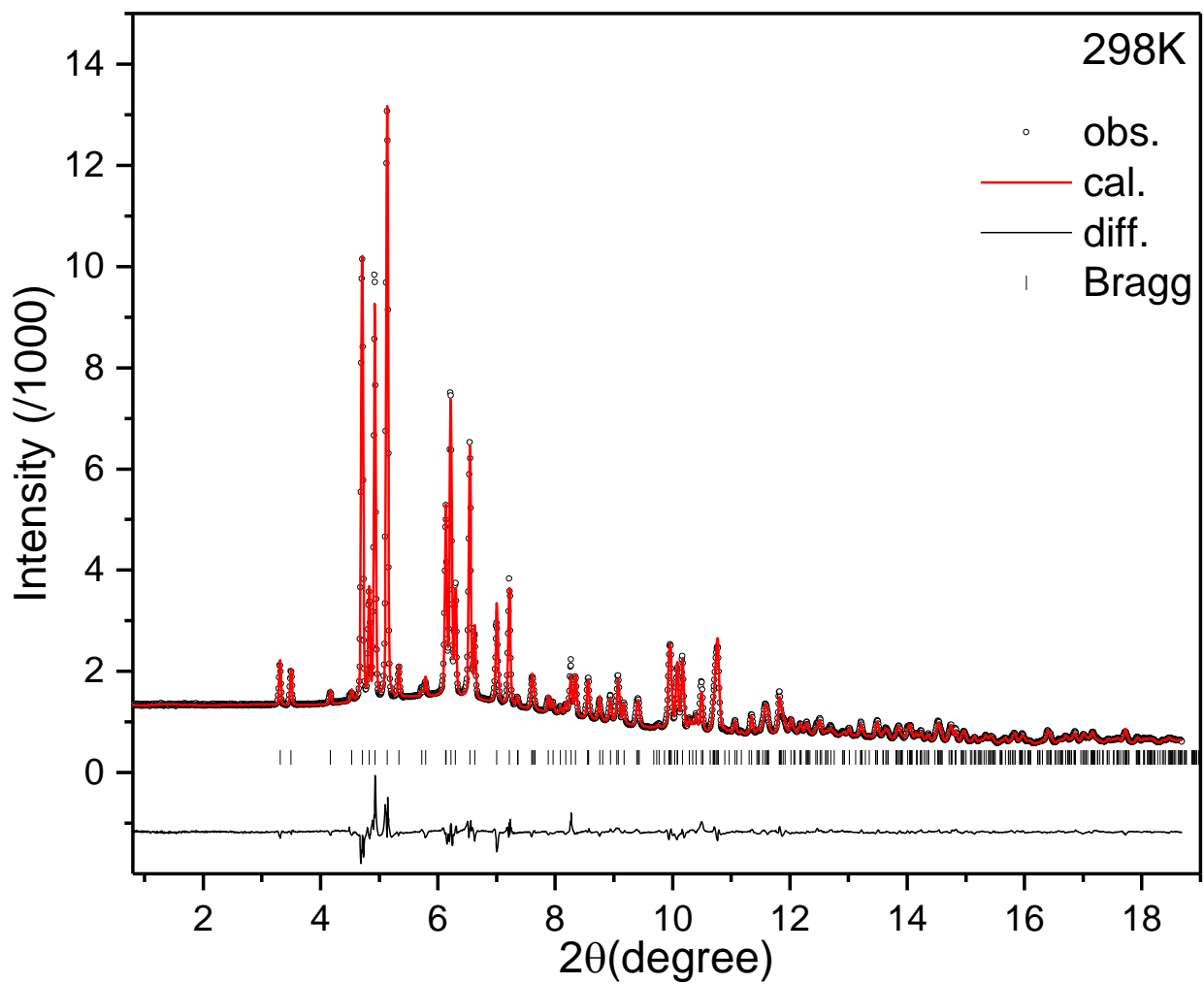

**Figure S8.** Experimental (circles), fitted (line), and difference (line below observed and calculated patterns) XRD profiles for K-7,9-$C_2B_9H_{12}$ at 298 K ($\lambda$=0.45399Å). Vertical bars indicate the calculated positions of Bragg peaks of K-7,9-$C_2B_9H_{12}$ (S.G. $P2_1/a$). $R_{wp}$=0.0318, $R_p$=0.0209, $\chi^2$=1.12. Refined lattice parameters of monoclinic phase K-7,9-$C_2B_9H_{12}$ at room temperature: $a$=10.7796(4)Å, $b$=11.4692(4) Å, $c$=7.43895(27) Å, $\beta$=93.5186(16)°, and V=917.97(8)Å$^3$.

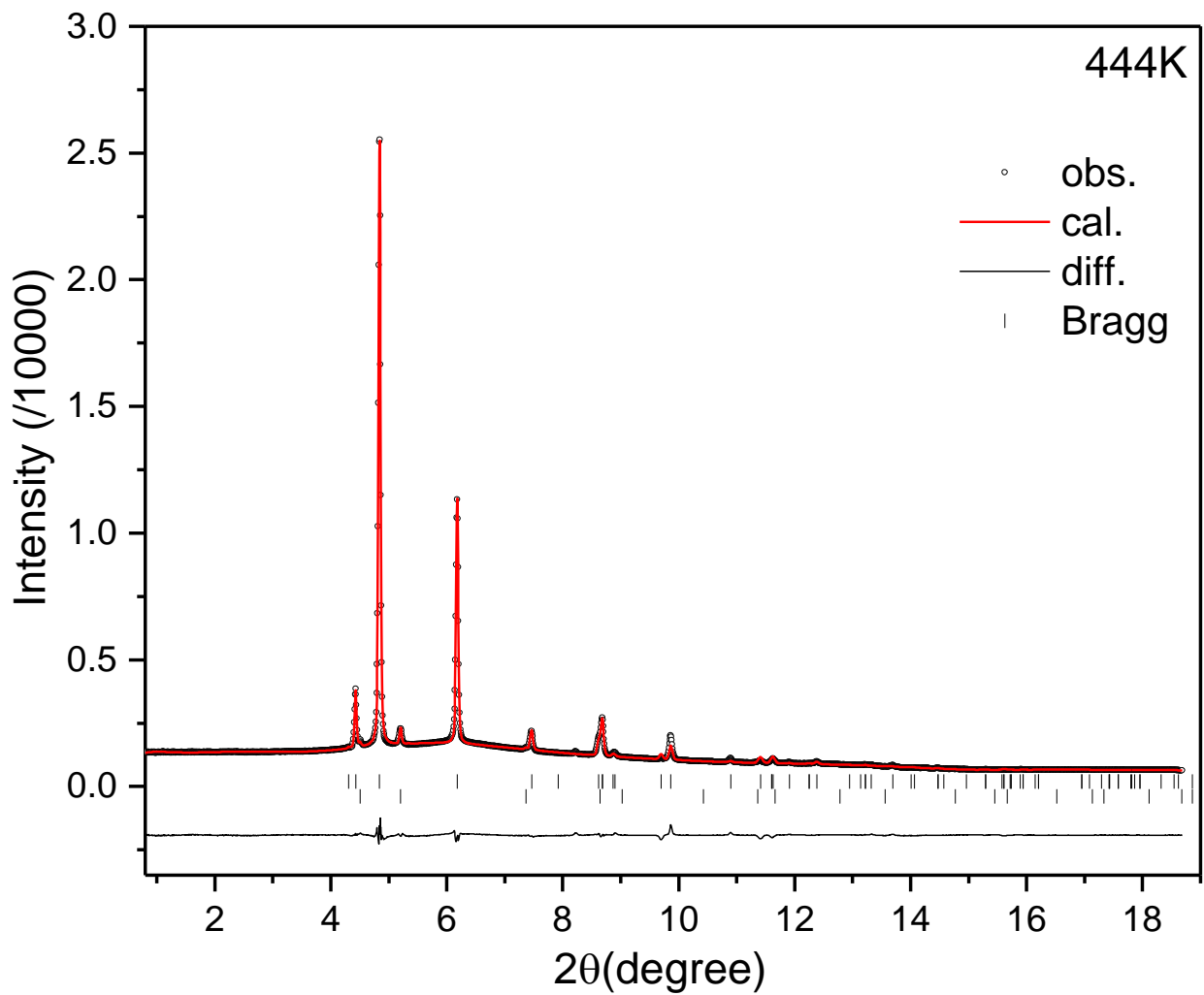

**Figure S9.** Experimental (circles), fitted (line), and difference (line below observed and calculated patterns) XRD profiles for K-7,9-$C_2B_9H_{12}$ at 444 K ($\lambda$=0.45399Å). Vertical bars indicate the calculated positions of Bragg peaks of K-7,9-$C_2B_9H_{12}$ (S.G. *P*-31*c*, wt Frac. 97.894(7) % and S.G. Fm-3m, wt Frac. 2.11(8)%) (from the top). $R_{wp}$=0.0242, $R_p$=0.0148, $\chi^2$=1.33. Refined lattice parameters of trigonal phase K-7,9-$C_2B_9H_{12}$ (*P*-31*c*) at 444K: *a*=6.9728(7)Å, *c*=11.7364(10)Å, and V=494.18(12)Å$^3$.

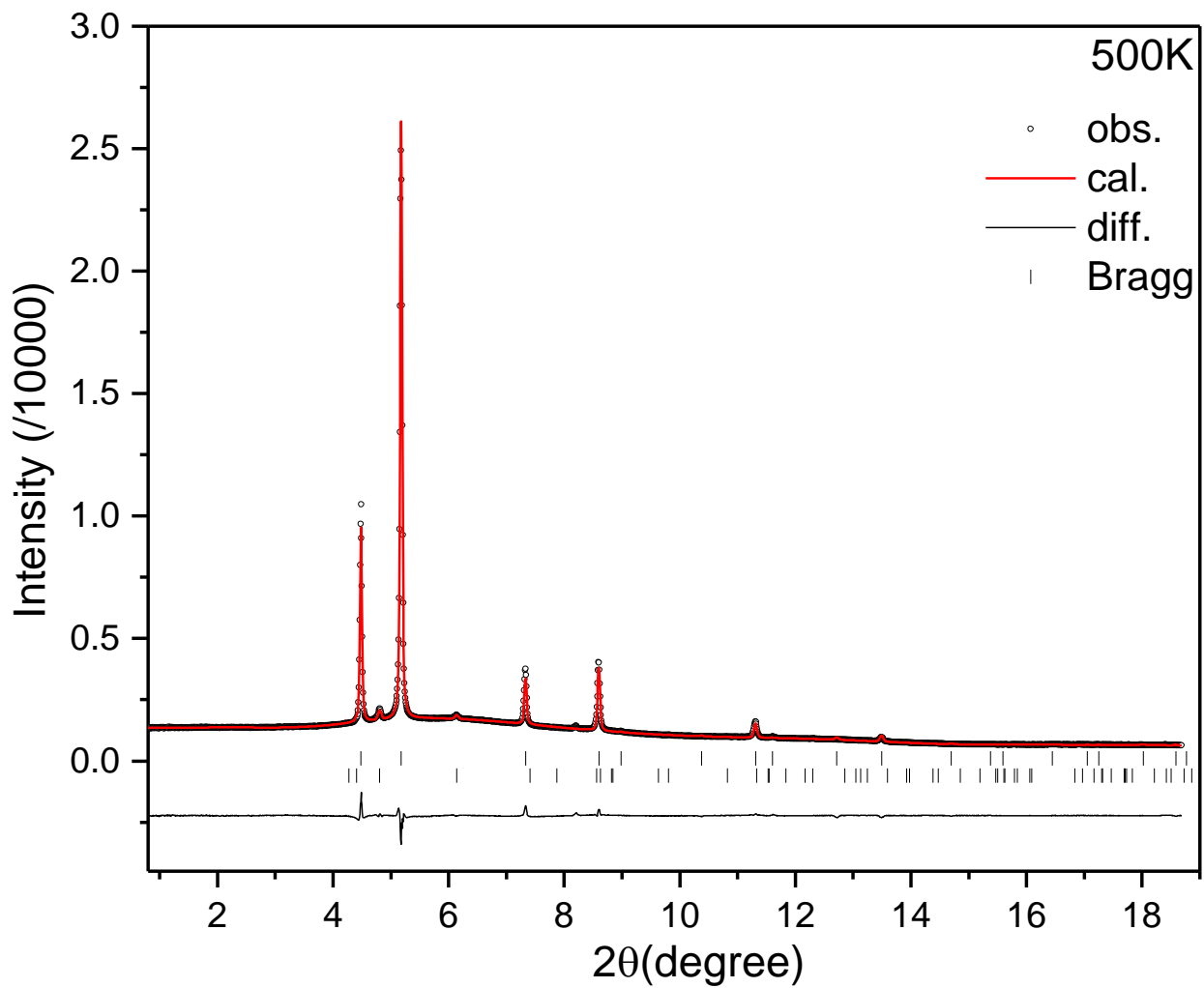

**Figure S10.** Experimental (circles), fitted (line), and difference (line below observed and calculated patterns) XRD profiles for K-7,9-$C_2B_9H_{12}$ at 500 K ($\lambda$=0.45399Å). Vertical bars indicate the calculated positions of Bragg peaks of K-7,9-$C_2B_9H_{12}$ (S.G. *Fm-3m*, wt Frac. 97.999(3) % and S.G. P-31c, wt Frac. 2.00(6) %) (from the top). $R_{wp}$=0.0209, $R_p$=0.0136, $\chi^2$=0.56. Refined lattice parameters of cubic phase K-7,9-$C_2B_9H_{12}$ (*Fm-3m*) at 500K are: *a*=10.0363(4) Å, V=1010.93(14) Å$^3$.

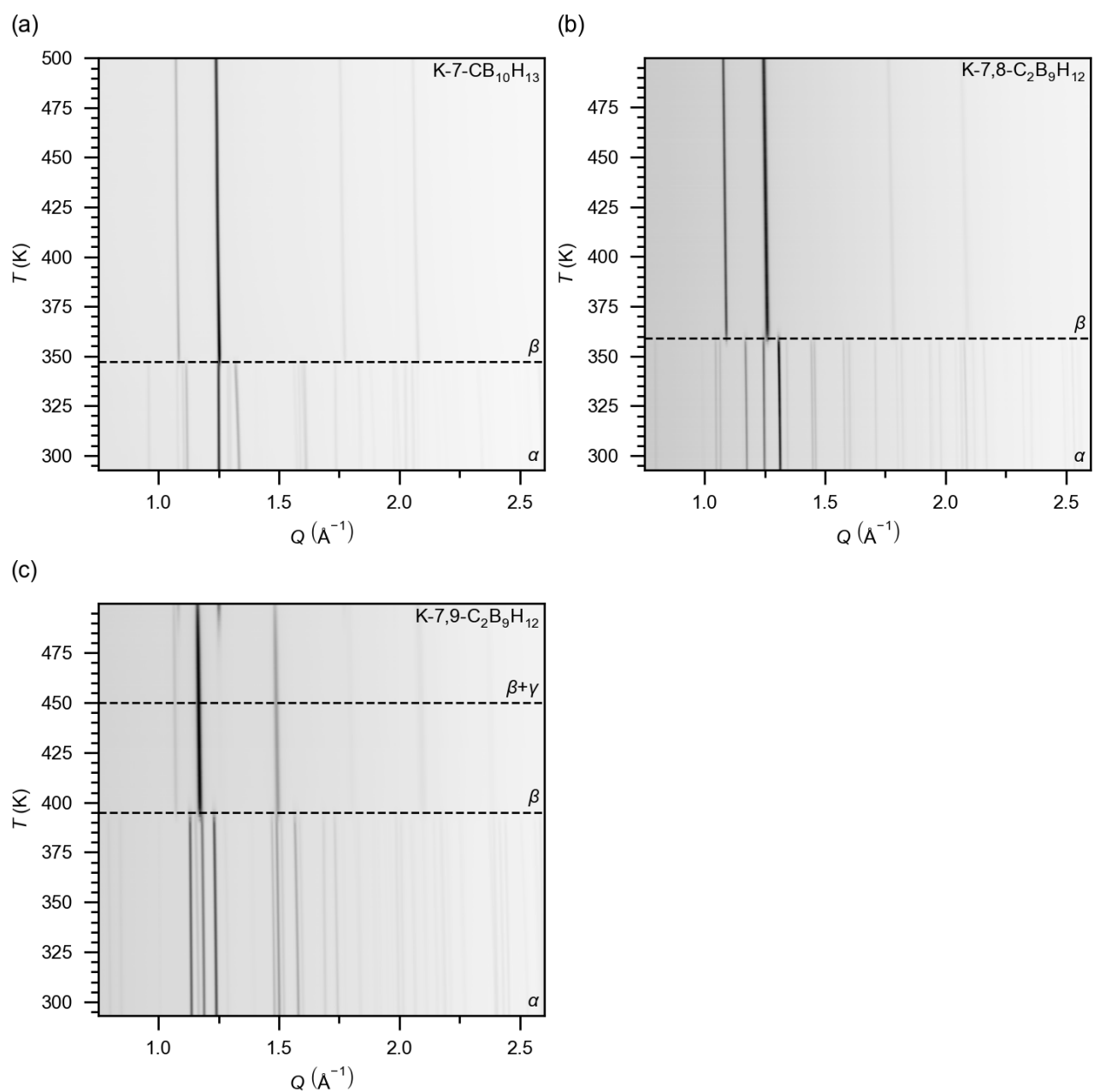

**Figure S11.** *In situ* synchrotron radiation powder X-ray diffraction data on a) K-7-CB$_{10}$H$_{13}$, b) K-7,8-C$_2$B$_9$H$_{12}$, and c) K-7,9-C$_2$B$_9$H$_{12}$. The dotted lines indicate phase transitions, and the phases present in the given temperature regions are shown with the respective Greek nomenclature.

**Quasielastic neutron scattering results**

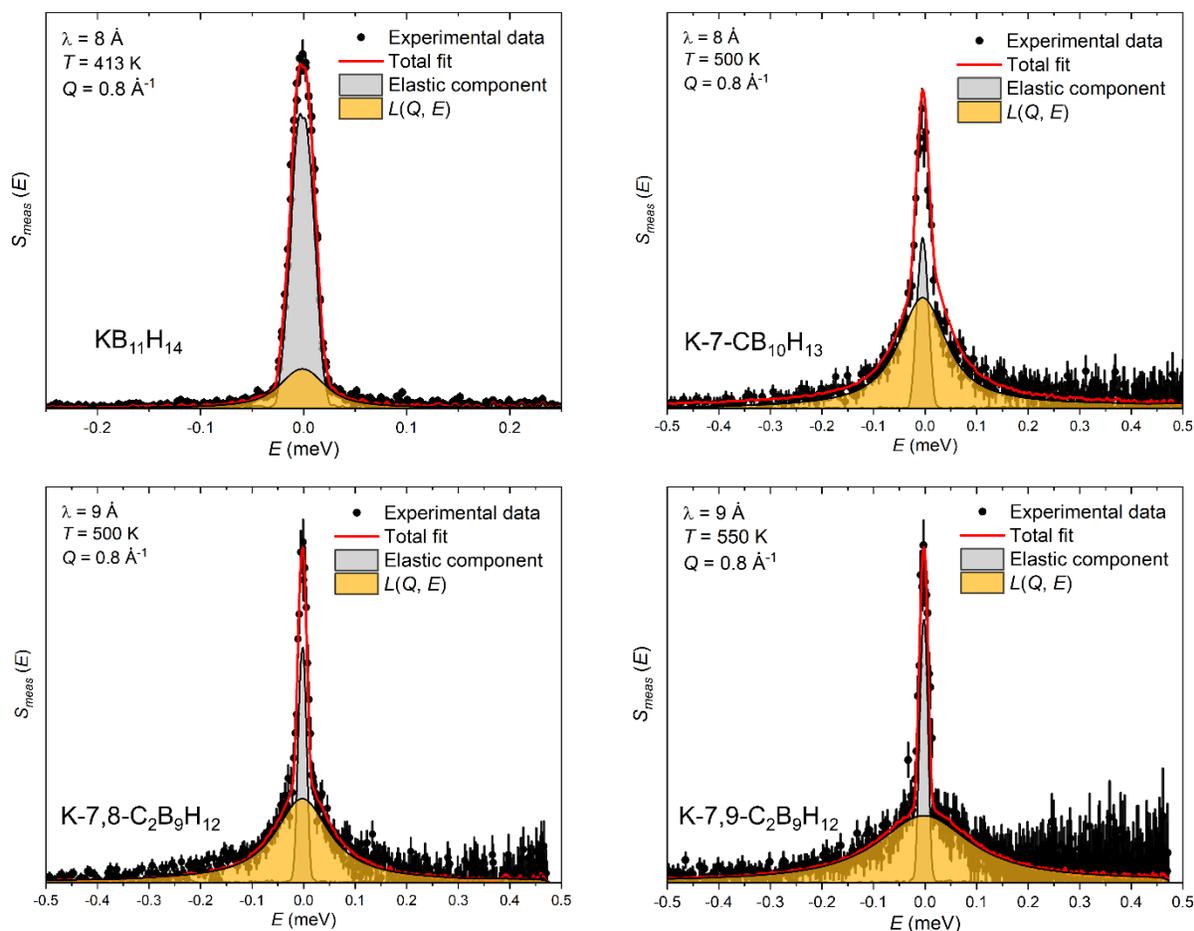

**Figure S12.** Exemplary QENS spectra for the four K-*nido*-compounds at Q = 0.8 Å$^{-1}$ for the designated temperatures and incident neutron wavelengths λ (either 8 Å with 30 μeV fwhm resolution or 9 Å with 22 μeV fwhm resolution). Spectra were fit with a delta function (elastic component) and one Lorentzian quasielastic component, both convoluted with the instrumental resolution function, on top of a flat background. N.B., the relatively larger elastic component for KB$_{11}$H$_{14}$ at 413 K reflects an increased fraction of immobile hydrogen atoms due to the presence of a substantial fraction of ordered α-phase.

Below are model expressions for the elastic incoherent structure factors (EISFs) associated with various reorientational mechanisms of *nido*-(carba)borate anions such as considered in Figure 7 of the text. The EISFs reflect the composite reorientational motions of all H atoms associated with the *nido*-(carba)borate anions. These expressions are constructed according to the general methodology described in Ref. S1. It should be noted that, although these undeca(carba)borate anions display somewhat minor distortions from an ideal icosahedral cluster, ignoring these distortions as well as assuming that the reorientational center is the icosahedral geometric center instead of the slightly shifted center of gravity of the lower-symmetry anions will simplify the various EISF expressions without significantly compromising their accuracy. Moreover, all jump distances of the $H_{exo}$ apical atoms were approximated from neutron powder diffraction (NPD) structural data for $Cs_2B_{12}D_{12}$ [S2], since the highly symmetric $B_{12}H_{12}^{2-}$ anion closely mimics the dimensions of the *nido*-anions, and NPD, when available, provides more accurate nuclear positions than XRD for H atoms. The various jump distances of all extra $H_{endo}$ atoms associated with the pentagonal aperture were assumed to be the same for all anions considered and were estimated from DFT structural optimizations of the isolated *nido*-anions, since these same optimized anion structures were used as rigid bodies in all the SXRD refinements.

Since the number of terminally bonded apical H atoms for the four *nido*-(carba)borate anions are identical (*i.e.*, eleven $H_{exo}$ atoms per anion), the small differences between EISFs for the different anions result from the differences in the number of extra endo H atoms (i.e., one $H_{endo}$ atom for the two $C_2B_9H_{12}^-$ isomers (7,8-$C_2B_9H_{12}^-$ and 7,9-$C_2B_9H_{12}^-$), two $H_{endo}$ atoms for 7-$CB_{10}H_{13}^-$, and three $H_{endo}$ atoms for $B_{11}H_{14}^-$). As stated in the text, for the sake of simplicity, the uniaxial axis was chosen to be the quasi-$C_5$-symmetric axis perpendicular to the open-nest pentagonal aperture of the *nido*-anions, although it can be shown that choosing another preferred axis leads only to minor changes in the corresponding EISF and only above $Q \approx 1.5$ Å$^{-1}$, as exemplified later below for 7-$CB_{10}H_{13}^-$.

For uniaxial five-fold ($2\pi/5$ radian) jumps around the chosen quasi-$C_5$-symmetric axis, one apical $H_{exo}$ atom remains stationary, while the other ten apical $H_{exo}$ atoms visit five angular jump positions with jump distances of $d_2 \approx 3.05$ Å and $d_4 \approx 4.93$ Å, and the additional $H_{endo}$ atoms visit five angular jump positions with relatively smaller jump distances of $r_2 \approx 1.14$ Å and $r_4 \approx 1.84$ Å. The resulting model EISFs for the three different types of anions are defined as:

$$\text{EISF}_{C5}(C_2B_9H_{12}^-) = \frac{1}{12} + \frac{10}{12} \times \frac{1}{5}[1 + 2j_0(Qd_2) + 2j_0(Qd_4)] + \frac{1}{12} \times \frac{1}{5}[1 + 2j_0(Qr_2) + 2j_0(Qr_4)], \quad \text{(S1)}$$

$$\text{EISF}_{C5}(\text{CB}_{10}\text{H}_{13}^{-}) = \tfrac{1}{13} + \tfrac{10}{13} \times \tfrac{1}{5}[1 + 2j_0(Qd_2) + 2j_0(Qd_4)] + \tfrac{2}{13} \times \tfrac{1}{5}[1 + 2j_0(Qr_2) + 2j_0(Qr_4)], \quad (S2)$$

$$\text{EISF}_{C5}(\text{B}_{11}\text{H}_{14}^{-}) = \tfrac{1}{14} + \tfrac{10}{14} \times \tfrac{1}{5}[1 + 2j_0(Qd_2) + 2j_0(Qd_4)] + \tfrac{3}{14} \times \tfrac{1}{5}[1 + 2j_0(Qr_2) + 2j_0(Qr_4)], \quad (S3)$$

where $j_0(x) = \sin(x)/x$ is the zeroth-order spherical Bessel function. We caution that the $\text{H}_{endo}$ contributions to $\text{EISF}_{C5}(\text{B}_{11}\text{H}_{14}^{-})$ may be oversimplified since the known fluxional nature of these three H atoms around the pentagonal aperture with both bridging and terminal bonding possibilities [S3] means that these atoms are possibly undergoing uniaxial rotational diffusion around the aperture ring while ten apical $\text{H}_{exo}$ atoms experience five-fold jumps, which would require a slight modification of the $\text{H}_{exo}$ portion of the EISF (see later below).

The EISF equations above will be altered in a minor fashion upon choosing a different symmetry axis of rotation. For example, for the 7-$\text{CB}_{10}\text{H}_{13}^{-}$ anion, choosing the quasi-$C_5$ rotation axis of the quasi-icosahedral framework as the one that contains the carbon apex (as considered in Ref. S4) results in two stationary $\text{H}_{exo}$ atoms instead of one, as well as different jump distances ($r_{2'} \approx 2.34$ Å and $r_{4'} \approx 3.79$ Å) for the two $\text{H}_{endo}$ atoms. In this case, the $\text{EISF}_{C5}$ is defined as:

$$\text{EISF}_{C5}(\text{CB}_{10}\text{H}_{13}^{-}) \approx \tfrac{2}{13} + \tfrac{9}{13} \times \tfrac{1}{5}[1 + 2j_0(Qd_2) + 2j_0(Qd_4)] + \tfrac{2}{13} \times \tfrac{1}{5}[1 + 2j_0(Qr_{2'}) + 2j_0(Qr_{4'})]. \quad (S4)$$

The resulting EISF curve is largely identical below $Q \approx 1.5$ Å$^{-1}$ to that resulting from Eq. S2 above, with a more noticeable deviation occurring only above $Q \approx 1.5$ Å$^{-1}$.

Uniaxial rotational diffusion (URD) around the quasi-$C_5$-symmetry axis used for Eqs. S1-S3 above can be approximated below $Q \approx 2.5$ Å$^{-1}$ by considering smaller-angle $\pi/5$ radian jumps yielding ten angular jump positions instead of five. In this case, one apical $\text{H}_{exo}$ atom still remains stationary, while the other ten apical $\text{H}_{exo}$ atoms visit ten positions with possible jump distances of $d_1 \approx 1.60$ Å, $d_2 \approx 3.05$ Å, $d_3 \approx 4.19$ Å, $d_4 \approx 4.93$ Å, and $d_5 \approx 5.18$ Å; and the additional $\text{H}_{endo}$ atoms visit ten positions with jump distances of $r_1 \approx 0.60$ Å, $r_2 \approx 1.14$ Å, $r_3 \approx 1.57$ Å, $r_4 \approx 1.84$ Å, and $r_5 \approx 1.94$ Å. The resulting model EISFs are defined as:

$$\text{EISF}_{\text{URD}}(C_2B_9H_{12}^-) \approx \frac{1}{12} + \frac{10}{12} \times \frac{1}{10}[1 + 2j_0(Qd_1) + 2j_0(Qd_2) + 2j_0(Qd_3) + 2j_0(Qd_4) + j_0(Qd_5)] +$$
$$\frac{1}{12} \times \frac{1}{10}[1 + 2j_0(Qr_1) + 2j_0(Qr_2) + 2j_0(Qr_3) + 2j_0(Qr_4) + j_0(Qr_5)], \quad (S5)$$

$$\text{EISF}_{\text{URD}}(CB_{10}H_{13}^-) \approx \frac{1}{13} + \frac{10}{13} \times \frac{1}{10}[1 + 2j_0(Qd_1) + 2j_0(Qd_2) + 2j_0(Qd_3) + 2j_0(Qd_4) + j_0(Qd_5)] +$$
$$\frac{2}{13} \times \frac{1}{10}[1 + 2j_0(Qr_1) + 2j_0(Qr_2) + 2j_0(Qr_3) + 2j_0(Qr_4) + j_0(Qr_5)], \quad (S6)$$

$$\text{EISF}_{\text{URD}}(B_{11}H_{14}^-) \approx \frac{1}{14} + \frac{10}{14} \times \frac{1}{10}[1 + 2j_0(Qd_1) + 2j_0(Qd_2) + 2j_0(Qd_3) + 2j_0(Qd_4) + j_0(Qd_5)] +$$
$$\frac{3}{14} \times \frac{1}{10}[1 + 2j_0(Qr_1) + 2j_0(Qr_2) + 2j_0(Qr_3) + 2j_0(Qr_4) + j_0(Qr_5)]. \quad (S7)$$

*N.B.*, the correction to Eq. S3 above for $\text{EISF}_{C5}(B_{11}H_{14}^-)$ to take into account the URD behavior of the fluxional $H_{endo}$ atoms easily follows from the formulation of Eq. S7 above. Hence, the modified Eq. S3 for uniaxial five-fold reorientations of the $B_{11}H_{14}^-$ anion becomes:

$$\text{EISF}_{C5}(B_{11}H_{14}^-) \approx \frac{1}{14} + \frac{10}{14} \times \frac{1}{5}[1 + 2j_0(Qd_2) + 2j_0(Qd_4)] +$$
$$\frac{3}{14} \times \frac{1}{10}[1 + 2j_0(Qr_1) + 2j_0(Qr_2) + 2j_0(Qr_3) + 2j_0(Qr_4) + j_0(Qr_5)]. \quad (S8)$$

Also, for uniaxial rotational diffusion of the $CB_{10}H_{13}^-$ anion around the axis containing the carbon apex (related to the expression for five-fold reorientations in Eq. S4), the five different jump distances for the two $H_{endo}$ atoms are $r_{1'} \approx 1.23$ Å $r_{2'} \approx 2.34$ Å, $r_{3'} \approx 3.22$ Å, $r_{4'} \approx 3.79$ Å, and $r_{5'} \approx 3.98$ Å; and the EISF becomes:

$$\text{EISF}_{\text{URD}}(\text{CB}_{10}\text{H}_{13}^-) \approx \tfrac{2}{13} + \tfrac{9}{13} \times \tfrac{1}{10}[1 + 2j_0(Qd_1) + 2j_0(Qd_2) + 2j_0(Qd_3) + 2j_0(Qd_4) + j_0(Qd_5)] +$$

$$\tfrac{2}{13} \times \tfrac{1}{10}\left[1 + 2j_0(Qr_{1'}) + 2j_0(Qr_{2'}) + 2j_0(Qr_{3'}) + 2j_0(Qr_{4'}) + j_0(Qr_{5'})\right].$$

(S9)

For multi-directional tumbling of the anions amongst the twelve positions of an icosahedron [S5], the EISF is composed of different contributions from the two types of H atoms. For the eleven $H_{\text{exo}}$ atoms, there are three different jump distances to reach the eleven other icosahedral positions: $d_{1'} \approx 3.05$ Å, $d_{2'} \approx 4.93$ Å, and $d_{3'} \approx 5.80$ Å. With the $H_{\text{exo}}$ atoms restricted to these particular icosahedral positions, the extra $H_{\text{endo}}$ atoms are capable of visiting a much larger and denser set of positions distributed over the surface of a sphere of radius $r_{\text{endo}} \approx 2.0$ Å. Hence, the EISF contributions of each of these atoms are best approximated by an isotropic rotational diffusion (IRD) model [$\text{EISF}_{\text{IRD}} = j_0^2(Qr_{\text{endo}})$; see Ref. S6]. The resulting composite model EISFs for the three different types of anions undergoing icosahedral tumbling are thus defined as:

$$\text{EISF}_{\text{icos}}(\text{C}_2\text{B}_9\text{H}_{12}^-) \approx \tfrac{11}{12} \times \tfrac{1}{12}\left[1 + 5j_0(Qd_{1'}) + 5j_0(Qd_{2'}) + j_0(Qd_{3'})\right] + \tfrac{1}{12}j_0^2(Qr_{\text{endo}}),$$

(S10)

$$\text{EISF}_{\text{icos}}(\text{CB}_{10}\text{H}_{13}^-) \approx \tfrac{11}{13} \times \tfrac{1}{12}\left[1 + 5j_0(Qd_{1'}) + 5j_0(Qd_{2'}) + j_0(Qd_{3'})\right] + \tfrac{2}{13}j_0^2(Qr_{\text{endo}}),$$

(S11)

$$\text{EISF}_{\text{icos}}(\text{B}_{11}\text{H}_{14}^-) \approx \tfrac{11}{14} \times \tfrac{1}{12}\left[1 + 5j_0(Qd_{1'}) + 5j_0(Qd_{2'}) + j_0(Qd_{3'})\right] + \tfrac{3}{14}j_0^2(Qr_{\text{endo}}). \quad (S12)$$

Finally, isotropic rotational diffusion of the anions consists of EISF contributions from the eleven $H_{\text{exo}}$ atoms and extra $H_{\text{endo}}$ atoms traversing spherical surfaces with respective radii $r_{\text{exo}} \approx 2.90$ Å and $r_{\text{endo}} \approx 2.0$ Å. The resulting model EISFs are defined as:

$$\text{EISF}_{\text{IRD}}(\text{C}_2\text{B}_9\text{H}_{12}^-) = \tfrac{11}{12}j_0^2(Qr_{\text{exo}}) + \tfrac{1}{12}j_0^2(Qr_{\text{endo}}), \quad (S13)$$

$$\text{EISF}_{\text{IRD}} (CB_{10}H_{13}^-) = \frac{11}{13}j_0^2(Qr_{\text{exo}}) + \frac{2}{13}j_0^2(Qr_{\text{endo}}), \tag{S14}$$

$$\text{EISF}_{\text{IRD}} (B_{11}H_{14}^-) = \frac{11}{14}j_0^2(Qr_{\text{exo}}) + \frac{3}{14}j_0^2(Qr_{\text{endo}}). \tag{S15}$$

Figure S13 depicts the various EISF curves described above. It should be noted that, similar to the relative behaviors of $\text{EISF}_{C5}$ and $\text{EISF}_{\text{URD}}$ curves, $\text{EISF}_{\text{icos}}$ and $\text{EISF}_{\text{IRD}}$ curves show little difference below $Q \approx 1.5$ Å$^{-1}$ and are also largely independent of the anion type.

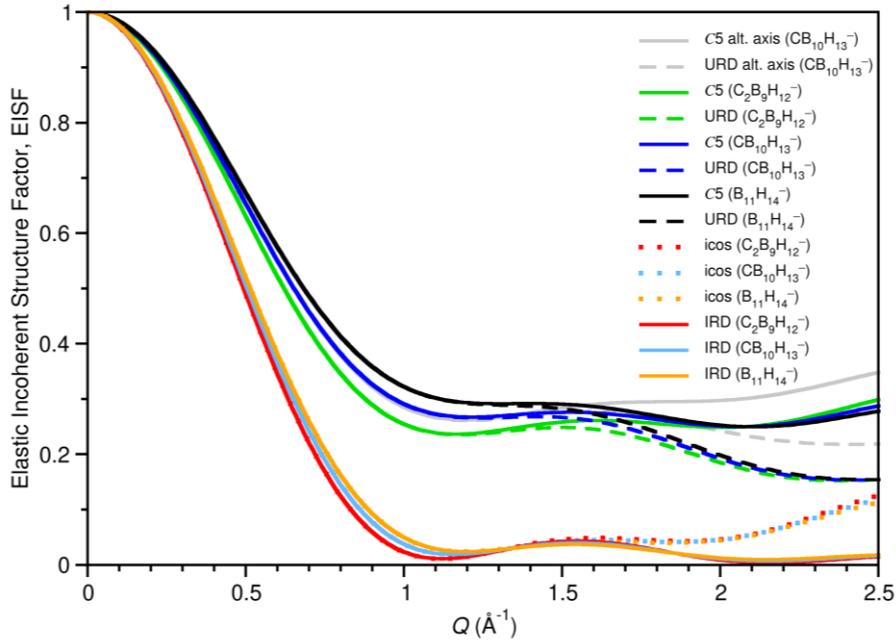

**Figure S13.** EISF model curves associated with various reorientation models for the different nido-anions. C5, URD, icos, and IRD refer to uniaxial five-fold jumps, uniaxial rotational diffusion, icosahedral tumbling, and isotropic rotational diffusion, respectively. All uniaxial reorientational curves assume that the reorientational axis is perpendicular to the open pentagonal aperture, except for "C5 and URD alt. axis ($CB_{10}H_{13}^-$)" (two gray curves representing Eqs. S4 and S9), which use a different reorientational axis running through the carbon apex as in Ref. S4. N.B., "C5 ($B_{11}H_{14}^-$)" (solid black curve) refers to Eq. S8 rather than Eq. S3.

**Ionic conductivity activation energy**

Ionic conductivity activation energies, $E_A$, were determined from linear fits of $\ln(\sigma T)$ vs. $1000/T$ plots of the ionic conductivity data and calculated with the formula

$$\sigma = \frac{\sigma_0}{T} \cdot e^{-\frac{E_A}{k_B \cdot T}}$$

where $\sigma$ is the ionic conductivity, $\sigma_0$ is the ionic conductivity prefactor, $T$ is the temperature, and $k_B$ is the Boltzmann constant.

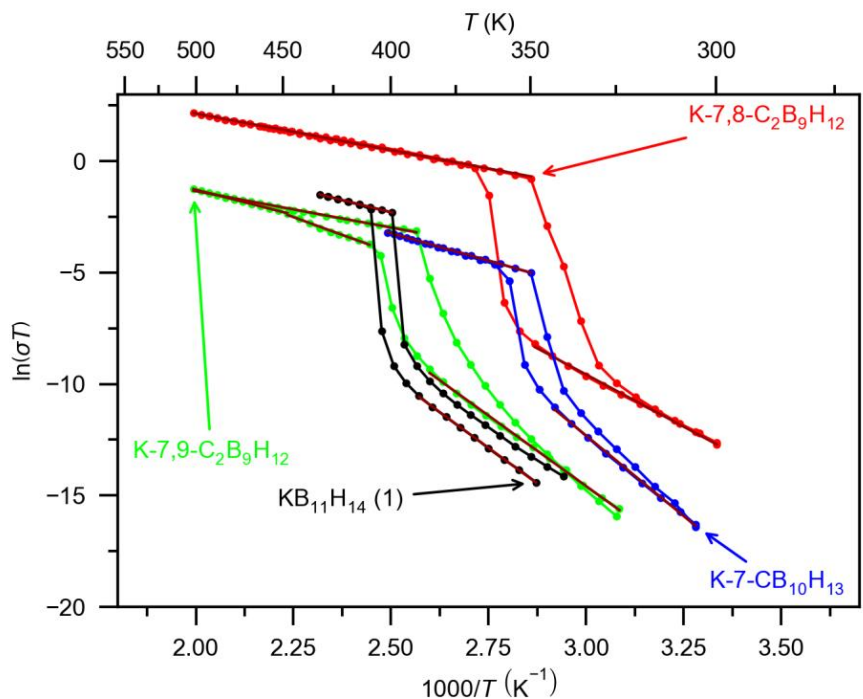

**Figure S14.** Plot of $\ln(\sigma T)$ vs. $1000/T$ data with the linear fits plotted as dark red lines.